\makeatletter
\def\@cons#1#2{\begingroup\let\@elt\relax\xdef#1{\ifx#1\relax\else#1\fi\@elt #2}\endgroup}
\makeatother

\RequirePackage{fix-cm}
\documentclass[final]{svjour3}                     
\smartqed  
\usepackage{graphicx}
\usepackage[colorlinks=true,allcolors=blue]{hyperref}
\usepackage{amsmath} 
\usepackage{mathtools} 
\usepackage{cleveref}
\usepackage{todonotes}
\presetkeys%
    {todonotes}%
    {inline}{}
\usepackage[round]{natbib}
\setcitestyle{aysep={}}

\usepackage{subcaption}
\usepackage{float}
\newcommand*{\diff}{\ensuremath{\mathop{}\!\mathrm{d}}}

\newcommand*{\od}[2]{\ensuremath{\frac{\diff #1}{\diff #2}}}
\renewcommand*{\v}[1]{\mathbf{#1}}

\newcommand*{\abs}[1]{\left\lvert#1\right\rvert}
\newcommand*{\sign}[1]{\text{sgn}\left(#1\right)}

\crefname{equation}{Eq.}{Eqs.}
\crefname{subsection}{Subsec.}{Subsecs.}
\crefname{section}{Sec.}{Secs.}
\crefname{table}{Tab.}{Tabs.}
\crefname{figure}{Fig.}{Figs.}
\crefname{subfigure}{Fig.}{Figs.}
\crefname{appendix}{}{}
\crefname{case}{case}{cases}

\begin{document}

\title{Rheology of Immiscible Two-phase Flow in Mixed Wet Porous Media: Dynamic Pore Network Model and Capillary Fiber Bundle Model Results
}
\titlerunning{Rheology of Immiscible Two-phase Flow in Mixed Wet Porous Media}        
\author{Hursanay~Fyhn     \and
        Santanu~Sinha     \and
        Subhadeep~Roy     \and
        Alex~Hansen}


\institute{Hursanay~Fyhn$^1$
          \quad\email{hursanay.fyhn@ntnu.com}\\            
          \and
          Santanu~Sinha$^{1,2}$ 
          \and
          Subhadeep~Roy$^1$     
          \and
          Alex~Hansen$^1$\\
          \at 1. PoreLab, Department of Physics, Norwegian University of Science and Technology (NTNU), NO-7491 Trondheim, Norway 
          \at 2. Beijing Computational Science Research Center, 10 East Xibeiwang Road, Haidian District, Beijing 100193, China
}

\date{Received: date / Accepted: date}

\maketitle

\begin{abstract}

Immiscible two-phase flow in porous media with mixed wet conditions was examined using a capillary fiber bundle model, which is analytically solvable, and a dynamic pore network model.
The mixed wettability was implemented in the models by allowing each tube or link to have a different wetting angle chosen randomly from a given distribution.
Both models showed that mixed wettability can have significant influence on the rheology in terms of the dependence of the global volumetric flow rate on the global pressure drop. 
In the capillary fiber bundle model, for small pressure drops when only a small fraction of the tubes were open, it was found that the volumetric flow rate depended on the excess pressure drop as a power law with an exponent equal to $3/2$ or $2$ depending on the minimum pressure drop necessary for flow. 
When all the tubes were open due to a high pressure drop, the volumetric flow rate depended linearly on the pressure drop, independent of the wettability.
In the transition region in between where most of the tubes opened, the volumetric flow depended more sensitively on the wetting angle distribution function and was in general not a simple power law.
The dynamic pore network model results also showed a linear dependence of the flow rate on the pressure drop when the pressure drop is large.
However, out of this limit the dynamic pore network model demonstrated a more complicated behavior that depended on the mixed wettability condition and the saturation.
In particular, the exponent relating volumetric flow rate to the excess pressure drop could take on values anywhere between $1.0$ and $1.8$. 
The values of the exponent were highest for saturations approaching $0.5$, also, the exponent generally increased when the difference in wettability of the two fluids were larger and when this difference was present for a larger fraction of the porous network.

\end{abstract}

\section{Introduction}
\label{introduction}

The study of rheology of two-phase flow in porous media is pivotal for many disciplines, and the wettability conditions of the system is an important factor that directly affects the rheology.
Examples for relevant disciplines include drug delivery in biology~\citep{vafai2010porous}, studies of human skin behavior relevant for cosmetic and medical sectors~\citep{elkhyat2001new}, creation of self-cleaning and fluid repelling materials relevant for textile industry~\citep{li2017review} and oil recovery~\citep{kovscek1993pore} and carbon dioxide sequestration~\citep{KREVOR2015221} in geophysics~\citep{blunt2017multiphase,marle1981multiphase}. 
All of these examples, dealing with different kinds of porous media, will benefit from a better understanding of two-phase flow under different wetting conditions.
Two-phase flow means simultaneous flow of two fluids in the same space.
When an immiscible fluid is injected into a porous medium filled with another fluid, different transient flow mechanisms occur depending on the flow conditions, such as capillary fingering~\citep{lenormand1989capillary}, viscous fingering~\citep{toussaint2005influence,maaloy1985viscous,lovoll2004growth} and stable displacement~\citep{frette1997immiscible,meheust2002interface}.
After the transient flow mechanisms have surpassed, steady state sets in, which is the regime in which the rheology of two-phase flow under different wetting conditions is examined in this work.  

Darcy's law is widely used to describe the flow of fluids through a porous medium which states that the volume of fluid flowing per unit area per unit time depends linearly on the applied pressure drop across a representative elementary volume in that porous medium~\citep{blunt2017multiphase}.
That is indeed the case for large applied pressures, however the linearity gets modified into a power law at the low pressure limit.
For the flow to start, the applied pressure has to overcome the disordered capillary barriers~\citep{sinha2012effective}. 
When the applied pressure is so small that it exceeds the capillary barriers in only parts of the porous medium, the capillary forces will be comparable to the viscous forces.
In this case, the volumetric flow rate scales nonlinearly with the pressure drop due to the fact that increasing the pressure drop by a small amount creates new connecting paths in addition to increase the flow in the previously connected paths.
Earlier works~\citep{roy2019effective,sinha2019dynamic,tallakstad2009steady,rassi2011nuclear,tallakstad2009steadyE,aursjo2014film,gaolin2020pore,zhang2021quantification} have provided experimental, theoretical and numerical evidences for this phenomena in porous media under uniform wetting conditions.
Instead of assuming uniform wetting conditions, we here investigate the same phenomena using non-uniform wetting conditions, theoretically and numerically.

The wetting condition of a porous medium is a major factor controlling the location, flow and distribution of fluids~\citep{anderson1986wettability}, and is a result of the interplay between the attractive forces on the surface of the adjoining materials. 
When two immiscible fluids flow in a porous medium, the relative values of the surface tensions between each pair of the three phases, namely the fluids and the solid, determine the wetting angle and hence the equilibrium configuration of the fluids.
In nature, the wettability of a porous medium tends to alter along the system and results in a range of different wetting angles. 
For instance, the internal surface of reservoir rocks is composed of many minerals with different surface chemistry and absorption properties, which can cause wettability variations~\citep{anderson1986wettability}.
There are different types of non-uniform wetting conditions depending on the degree of non-uniformity as well as the geometrical and topological distribution of regions with different wettability.
The examples include fractional wettability where grains with same type of wettability are packed together in different proportions or mixed wettability where there are continuous paths with one type of wettability~\citep{anderson1986wettability,salathiel1972oil}.
It is often useful to make these distinctions because the physical processes which create non-uniform wetting conditions can result in different forms of connectedness.
In this work, we want to study how the deviation from uniform wetting conditions affect the rheology.
Hence, it is desirable to isolate the effect of non-uniform wettability in terms of the mean wetting angle and the spread of the wetting angles.
To this end, we use mathematical models with wetting angles determined from various distributions.
We use the term \emph{mixed wet} to denote the resulting non-uniform wetting conditions, but note that this term can also imply geometrical effects mentioned above which are not considered here.
We leave for future work the problem of how others types of non-uniform wetting conditions can affect the rheology further.
A mechanism for a correlated wettability distribution for pore-network modeling, where the wettability depends on the connected oil paths, was demonstrated previously by some of the authors of this manuscript \citep{flovikSinha2015dynamic} and may be adopted in future.

Several works in the past have investigated multiphase flow in mixed wet porous media, and discovered clear discrepancies in the fluid behavior in uniform wet systems and in mixed wet systems.
Experimental studies have found that the main determinant of the filling sequence in a porous medium is the wettability rather than the pore size~\citep{scanziani2020dynamics,gao2020pore}.
There were also findings from experimental studies indicating that the processes where it is necessary to allow the flow of both fluids favor mixed wetting conditions~\citep{alratrout2018wettability,alhammadi2017situ}, such as oil recovery or fluid transport through membranes or in biological tissue. 
These experimental findings show the importance of understanding the effect of wettability even further, which is easier to do through analytical and numerical studies where large range of wetting conditions can be examined in short time.
In the papers by \cite{sinhaGrova2011local} and \cite{flovikSinha2015dynamic}, pore network models similar to the one used in the present article were used to investigate the effect of wettability alteration due to changes in salinity in oil-brine mixtures.
The wettability alterations were done by changing between either complete wetting and complete non-wetting conditions in the first article~\citep{sinhaGrova2011local}, and by changing the wetting angles continuously between two limits depending on the cumulative flow of the wetting phase in the second article~\citep{flovikSinha2015dynamic}.
The results from both show that local alterations of the wettability introduce qualitative changes in the flow patterns by destabilizing the trapped clusters.
While such past numerical studies provide important insight into the behaviors of mixed wet porous media and support the usefulness of mixed wettability, they consider limited cases of the wetting angle conditions and do not consider the effect of the applied pressure on the flow.
In the present work, we conduct a systematic analysis of the effect of mixed wetting conditions, both in terms of a wide range of different mean wetting angles as well as different spread of the wetting angles.
In doing so we manage to perform a direct study of the relation between the total volumetric flow rate and the pressure drop across the system as influenced by the mixed wettability. 

Stated more in detail, the investigations in this work have been carried out by, firstly, calculating the total volumetric flow rate in a model consisting of a bundle of capillary tubes with mixed wet properties~\citep{roy2019effective,sinha2013effective}.
Thereafter, case studies with various specific wetting angle distribution have been performed through numerical calculations which confirmed the analytical results in addition to providing a holistic picture.
Secondly, mixed wetting conditions have been implemented into a dynamic network model~\citep{sinha2019dynamic} where the motion of the fluid interfaces are followed through the porous medium.
The results confirm that the volumetric flow rate $Q$ indeed depends on the applied pressure drop $\Delta P$ as
\begin{equation}
Q\propto \left(|\Delta P|-|P_t|\right)^\beta\;,
\end{equation}
where $P_t$ is the minimum pressure drop necessary for flow.
The exponent $\beta>1$ in the low pressure limit and $\beta=1$ in the high pressure limit.
More specifically in the low pressure limit, the capillary fiber bundle model considering a simple system gives $\beta=2$ and $\beta=3/2$, while the dynamic pore network model considering a more sophisticated system gives values varying anywhere between $\beta\in[1.0,1.8]$ depending on the system wettability configuration.

The models and the wetting condition implementing methods, as well as previously existing relevant theories, are explained in \cref{sec:methodology}.
The theoretical and numerical results are presented and discussed in \cref{sec:results} and a conclusion summarizing the findings is given in \cref{sec:conclusion}.

\section{Methodology}%
\label{sec:methodology}

\subsection{The Capillary Fiber Bundle Model Description}%
\label{sub:the_capillary_fiber_bundle_model_description}

\begin{figure}[ht]
  \centering
  \includegraphics[width=0.47\linewidth]{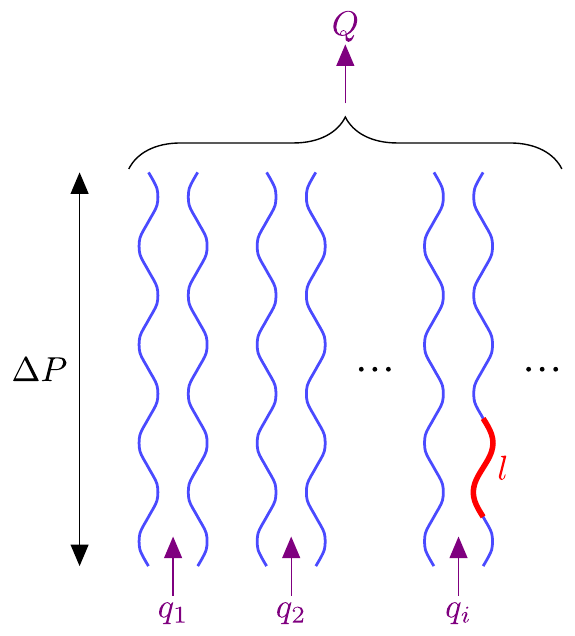}
  \caption{Capillary fiber bundle (CFB) model consists of a bundle of parallel sinusoidally shaped capillary tubes with period $l$.
  A global pressure drop $\Delta P$ drives the flow with volumetric flow rate $q_i$ in each tube $i$ which combine to produces a total volumetric flow rate $Q$.
}%
  \label{fig:CFB_sketch}
\end{figure}

The first model that is used to investigate immiscible two phase flow in mixed wet porous media is a capillary fiber bundle (CFB) model~\citep{roy2019effective,sinha2013effective}.
This model consists of a bundle of parallel capillary tubes, disconnected from each other, each carrying the two immiscible fluids. 
A typical porous medium normally has a varying radius for the links in the system, which is a factor that contributes to the capillary pressure being position dependent.
To emulate this effect, sinusoidal shaped tubes have been used in this model.
A sketch of the model is shown in \cref{fig:CFB_sketch}. 

As the main goal of this work is to examine the effect of mixed wettability, each one of the tubes in the CFB model has been given a wetting angle $\theta$ chosen randomly from a certain predefined distribution $\rho(\theta)$.
This means that each tube has the same assigned wetting angle over its entire length. 
The flow is driven by applying a global pressure drop $\Delta P$ over the system.
The total global volumetric flow rate $Q$ of the bundle of tubes is then calculated by considering the contributions to the flow given by each tube.
This calculation has been carried out both analytically and numerically.

\subsection{The Dynamic Pore Network Model Description}%
\label{sub:the_dynamic_pore_network_model_description}

\begin{figure}[ht]
  \centering
  \includegraphics[width=0.6\linewidth]{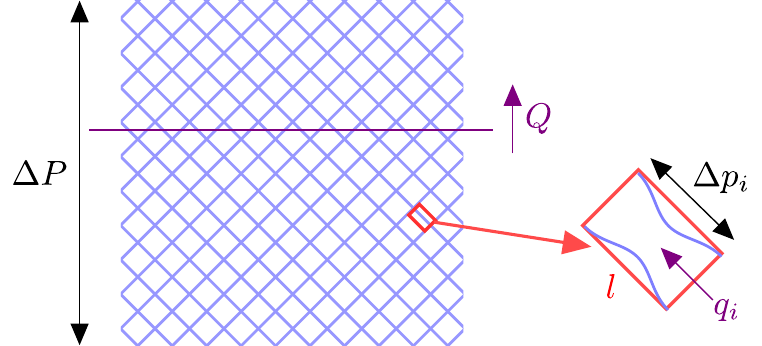}
  \caption{Two-dimensional dynamic pore network (DPN) model consists of hourglass shaped links with same length $l$ oriented with the same angle $(45^{\circ})$ from the flow direction.
  A global pressure drop $\Delta P$ drives the flow.
The total volumetric flow rate $Q$ is constant over all the cross sections normal to the direction of the overall flow. 
Each link $i$ has a local flow rate $q_i$, length $l$ and a pressure drop $\Delta p_i$ over it.}%
  \label{fig:DPN_sketch}
\end{figure}

The second model that is used for the investigations in this work is a dynamic pore network (DPN) model, a complex numerical model which is not analytically solvable \citep{sinha2019dynamic,amhb98,kah02,toh12,gvkh18}. 
A sketch of the network used in the model is given in \cref{fig:DPN_sketch} and a short description will be given here.
In this two-dimensional (2D) simulation, a porous network is modeled through a combination of links oriented with the same angle ($45^\circ$) from the flow direction and nodes connecting those links.
The movement of the two immiscible fluids are modeled through tracking of their interfaces at each instant in time. 
The fluids get distributed to the neighboring links when they reach a node at the end of the link in which they have been traveling.
The nodes themselves retain no volume.
Embracing the concept of varying radius of the typical porous media, similar to what has been done in the CFB model, the links in this model is made to be hourglass shaped as shown with a zoomed in sketch in \cref{fig:DPN_sketch}. 

The mixed wettability has been implemented into the DPN model by randomly choosing a wetting angle $\theta$ for each one of the links in the network from a certain predefined wetting angle distribution $\rho(\theta)$.
As in the case with the first model, the flow in the DPN model is driven by a pressure drop $\Delta P$ over the system.
When using periodic boundary conditions, $\Delta P$ is defined across a period of the system.
The total volumetric flow rate $Q$ is constant over all the cross sections normal to the direction of the overall flow, as the one illustrated with a horizontal line in \cref{fig:DPN_sketch}. 

\subsection{Commonalities}%
\label{sub:commonalities}

There are several commonalities in the two models. 
The smallest computational unit, which will be denoted SCU for ease of reference, in the CFB model is a single tube and that in the DPN model is a single link.
Even though each SCU in the two models has uniform wettability, the entire system consisting of various such entities together describes a mixed wet porous media.
In both models, the radius of each SCU $i$, with cylindrical symmetry around the center axis $\v x$ has the form
\begin{equation}
  \label{eq:radius}
  r_i(x) = \frac{r_{0,i}}{1-a\cos\left( \frac{2\pi x}{l} \right)} , 
\end{equation}
where $r_{0,i}$ is a constant and $a$ is the amplitude of the periodic variation.
In the DPN model, $l$ in \cref{eq:radius} is the length of the links, since the shape of the links covers only one period of oscillation, giving an hourglass form as shown in \cref{fig:DPN_sketch}.
In the CFB model, $l$ in \cref{eq:radius} is the wavelength of the shape of the tubes, as shown in \cref{fig:CFB_sketch}.

The flow within SCUs are governed by the following equations.
In SCU $i$, the capillary pressure $p_{c,i}(x)$ across an interface between the two immiscible fluids with wetting angle $\theta_i$ can be derived from the Young-Laplace equation to be~\citep{blunt2017multiphase}
\begin{equation}
  \label{eq:pc}
  p_{c,i}(x) = \frac{2\sigma \cos{\theta_i}}{r_i(x)} ,
\end{equation}
where $\sigma$ is the surface tension.
For each SCU with length $l'$ experiencing a pressure drop $\Delta p$ across its body, the fluid within it is forced to move due to the force exerted by the total effective pressure.
Total effective pressure is the difference between $\Delta p$ and the total capillary pressure $\sum_k p_c(x_k)$ due to all the interfaces with positions $x_k \in [0,l']$.
Assuming that the radius does not deviate too much from its average value $\bar r_i$, the volumetric flow rate $q_i$ in SCU $i$ is given by~\citep{sinha2019dynamic,washburn1921dynamics}
\begin{equation}
  \label{eq:qi}
  q_i =
  - \frac{\pi \bar r_i ^4}{8\mu_i l'} \left(\Delta p_i - \sum_k p_{c,i}(x_k)\right) 
\end{equation}
where $\mu_i$ is the saturation weighted viscosity of the fluids given by 
\begin{equation}
  \label{eq:mu}
  \mu_i = s_{A,i}\mu_{A} + s_{B,i}\mu_{B}.
\end{equation}
Here, $s_{A,i}=l'_{A,i} / l'$ and $s_{B,i}=l'_{B,i} / l'$ are saturations of the two fluids $A$ and $B$ with viscosities $\mu_{A}$ and $\mu_{B}$ and lengths $l'_{A,i}$ and $l'_{B,i}$.
In the DPN model, $l'$ is the same as $l$ from \cref{eq:radius}.
In the CFB model, $l'$ is the length of the whole tube.
The capillary number $\mathit{Ca}$, which is the ratio of viscous to capillary forces, is related to $q_i$ through $\mathit{Ca}=\mu Q/ (\sigma \alpha)$ where $Q$ is the sum of all $q_i$ through a cross sectional area $\alpha$~\citep{sinha2019dynamic}.

\begin{figure}[H]
  \centering
  \includegraphics[width=1.0\linewidth]{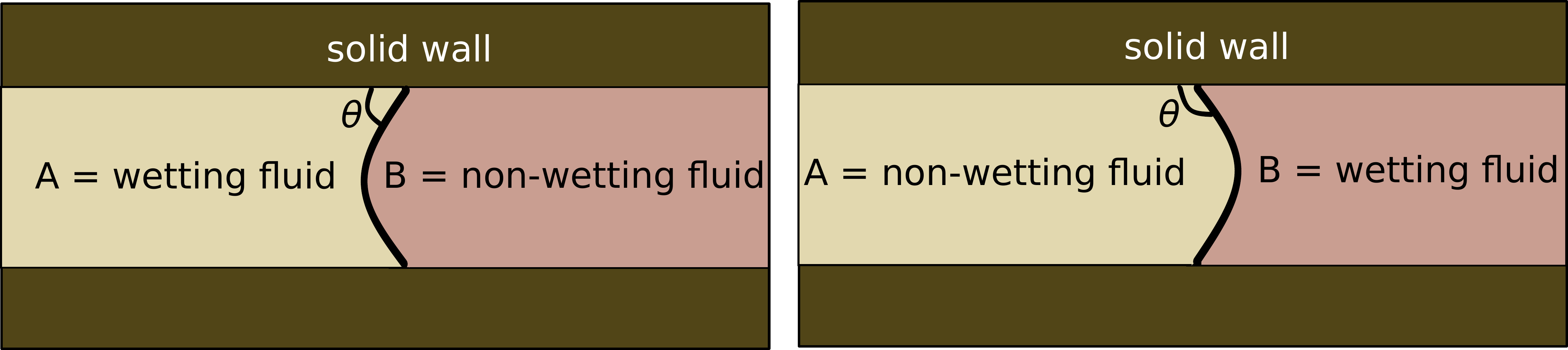}
  \caption{The wetting angles $\theta$ in this work are defined through fluid $A$. 
The fluid that makes the smallest angle with the solid wall is the wetting fluid while the other one is the non-wetting fluid.}%
  \label{fig:wettabilityflow}
\end{figure}

Note that due to the incompressible nature of the fluids examined in this work, $q$ given by \cref{eq:qi} is the same for any position along a single SCU.  
Also note that all $\theta$ in this work are defined through fluid $A$, as shown in \cref{fig:wettabilityflow}, which means the wetting angles of fluid $B$ are $180^\circ-\theta$.
The fluid that makes the smallest angle with the solid wall is the wetting fluid in that region of the pore space and the other fluid is the non-wetting fluid.

\section{Results}%
\label{sec:results}

\subsection{The Capillary Fiber Bundle Model Results}%
\label{sub:the_capillary_fiber_bundle_model_results}

The analysis of flow in a single tube is presented in \cref{ssub:a_single_tube}.
The theory is extended to a bundle of tubes with non-uniform wettability in \cref{ssub:a_bundle_of_tubes}.
The results from numerically solving the equations derived in \cref{ssub:a_bundle_of_tubes} give a holistic picture of the system.
They are presented in \cref{ssub:numerical_results} and agree with the theoretical calculations from \cref{ssub:a_bundle_of_tubes}.
A further explanation of the results is given in \cref{ssub:the_origin_of_beta}.

\subsubsection{A Single Tube}%
\label{ssub:a_single_tube}

In the paper by \cite{sinha2013effective}, they calculate the flow properties in a capillary tube with $\cos(\theta) = 1$.
Here, for a single tube, we will follow their calculations while keeping $\theta$ as a variable as it is needed for the rest of the work.
The parameters $r_0$, $\mu$, $l$, $\sigma$, $a$, $l'$ and $\Delta p$, as given in \cref{eq:radius,eq:pc,eq:qi,eq:mu}, are kept constant for all the tubes.
All the tubes have the same length $l'= L$ and global applied pressure $\Delta p=\Delta P$.

We start by considering a capillary tube with $N$ bubbles.
A ``bubble'' is one type of fluid restricted on two sides by the other fluid.
Each bubble $j$ has the center of mass position $x_j$ and a width $\Delta x_j$.
From \cref{eq:radius,eq:pc,eq:qi,eq:mu} we find that the volumetric flow rate through one tube is
\begin{equation}
  \label{eq:qiCFB}
  q = - \frac{\pi \bar r ^4}{8\mu L} 
  \left[ \Delta P + 
    \sum_{j=0}^{N-1}  \cos{\theta}
  \frac{4a\sigma}{r_0} \sin\left( \frac{\pi \Delta x_j}{l}\right)
    \sin\left( \frac{2\pi }{l} (x_0 + \delta x_j)\right) 
  \right],
\end{equation}
where $\delta x_j = x_j-x_0$.
Due to the incompressible nature of the fluids, the velocity of the bubbles is approximately constant along the axis of flow and equal to $\diff x_0 / \diff t \approx \diff x_j / \diff t \approx q/(\pi \bar r^2)$.
In addition, the effect of the variation in $\Delta x_j$ and $\delta x_j$ can be assumed to be small.
With this, \cref{eq:qiCFB} can be rewritten as
\begin{equation}
  \label{eq:dxbdt}
  \od{x_0}{t} =
  - \frac{\bar r ^2}{8\mu L} 
  \left[ \Delta P + 
\cos{\theta}
\left(
    \Gamma_s \sin\left( \frac{2\pi x_0}{l} \right)
    +
    \Gamma_c \cos\left( \frac{2\pi x_0}{l} \right)
    \right)
  \right],
\end{equation}
where
\begin{equation}
  \Gamma_s =
    \sum_{j=0}^{N-1} \frac{4a\sigma}{r_0} \sin\left( \frac{\pi \Delta x_j}{l}\right) 
    \cos\left( \frac{2\pi \delta x_j}{l}\right) 
\end{equation}
and 
\begin{equation}
  \Gamma_c = 
    \sum_{j=0}^{N-1} 
  \frac{4a\sigma}{r_0} \sin\left( \frac{\pi \Delta x_j}{l}\right)
 \sin\left( \frac{2\pi \delta x_j}{l}\right) .
\end{equation}
With algebraic manipulations, \cref{eq:dxbdt} can be rewritten as
\begin{equation}
  \label{eq:dxdt}
  \od{x}{t} =
  - \frac{\bar r ^2}{8\mu L} 
  \left[ \Delta P - 
  \cos{\theta}
  \sqrt{\Gamma_s^2 + \Gamma_c^2}
  \sin\left( \frac{2\pi x}{l} \right)
  \right],
\end{equation}
where $x = x_0+[\arctan(\Gamma_c/\Gamma_s) + \pi]l/2\pi$.
Defining 
\begin{equation}
  \label{eq:gamma}
  \gamma = k_\gamma \cos\theta
\end{equation}
with
\begin{equation}
  \label{eq:kgamma}
  k_\gamma = 
  \sqrt{\Gamma_s^2 + \Gamma_c^2}
  ,
\end{equation}
write \cref{eq:dxdt} as
\begin{equation}
  \label{eq:dxdtwithgamma}
  \od{x}{t} =
  - \frac{\bar r ^2}{8\mu L} 
  \left[ \Delta P - 
    \gamma
  \sin\left( \frac{2\pi x}{l} \right)
  \right].
\end{equation}

We wish to calculate the average velocity of the bubbles as they travel from one end to the other end of a tube segment with length $l$ using a time $T$,
\begin{equation}
  \label{eq:avdxbdt}
  \left\langle \od{x}{t} \right\rangle
  = \frac{l}{T}. 
\end{equation}
$T$ can be calculated by using the equation of motion in \cref{eq:dxdtwithgamma},
\begin{align}
  T 
  \nonumber
  &= \int_0^l \left(\od{x}{t}\right)^{-1} \diff x \\
  \nonumber
  &= - \frac{8\mu L}{\gamma\bar r^2} \int_0^l \frac{1}{ \frac{\Delta P}{\gamma} -\sin \left( \frac{2\pi x}{l} \right) }  \diff x \\
  \label{eq:T}
  &= -\frac{8\mu Ll}{\bar r^2} \frac{\sign{\Delta P}}{\sqrt{\Delta P^2 -\gamma^2}} 
  .
\end{align}
Inserting the result in \cref{eq:T} into \cref{eq:avdxbdt} and using the relation $\diff x / \diff t \approx q/(\pi \bar r^2)$ gives the average volumetric flux equation
\begin{equation}
  \label{eq:avq}
  \langle q \rangle =
  -\frac{\pi\bar r^4}{8\mu L} \sign{\Delta P} \Theta(\abs{\Delta P} -\abs{\gamma}) \sqrt{\Delta P^2 -\gamma^2} ,
\end{equation}
where $\Theta$ is the Heaviside step function.
From \cref{eq:avq} we see that, on average, the direction of flow is opposite to the pressure drop, as expected.
Additionally, we see that for a nonzero flow, $\Delta P$ needs to exceed a certain threshold $\gamma$ that is specific for the tube.


\subsubsection{A Bundle of Tubes}%
\label{ssub:a_bundle_of_tubes}

In the CFB model, the global volumetric flow rate $Q$ of a bundle of tubes is the sum of the time-averaged individual volumetric flow rates $\langle q\rangle$ of all the tubes that carry flow.
As remarked at the end of \cref{ssub:a_single_tube}, the tubes that carry flow are those that have a threshold $\gamma$ that satisfies the requirement $\abs{\Delta P}>\abs{\gamma}$. 
We will define a quantity $P_t$ which is the minimum possible $\gamma$ a tube can have.
This means that the first active path across the entire system occurs once $\Delta P$ exceeds $P_t$.
Let us also define $\gamma_\text{max}$ as the maximum possible $\gamma$ a tube can have, for later use.
The factors that determine $\gamma$ can be seen from \cref{eq:gamma}. 
Among those, $\theta$ is the only variable that varies from tube to tube, while the other quantities are set to be universal.
Under a constant $\Delta P$, what determines which tubes in the bundle will conduct flow, while others do not is therefore their $\theta$.
Using \cref{eq:gamma} and that $\theta\in [0^{\circ},180^{\circ}]$, the requirement for flow to happen in a tube can be rewritten as
\begin{equation}
  \label{eq:flowrequirement}
  \theta_a <\theta< \theta_{P_t}^+
  \quad \text{and} \quad
  \theta_{P_t}^- <\theta< \theta_b,
\end{equation}
with
\begin{align}
  \label{eq:thetaA}
  &\theta_a = \arccos \left( \text{min} \left( 1,\frac{\abs {\Delta P}}{k_\gamma} \right) \right),\\
  \label{eq:thetaB}
  &\theta_b = \arccos \left( \text{max} \left( -1,-\frac{\abs {\Delta P}}{k_\gamma} \right) \right) = 180^\circ - \theta_a,\\
  \label{eq:thetaGamma}
  &\theta_{P_t}^\pm = \arccos \left( \pm\frac{\abs {P_t}}{k_\gamma}  \right)
  .
\end{align}
Note that $\theta_{P_t}^+ = \theta_{P_t}^-$ when $\abs{P_t}=0$.

Since the flow requirement in \cref{eq:flowrequirement} indicates the range of $\theta$ that give the system a nonzero flow, $Q$ can be expressed as a function of the probability distribution $\rho(\theta)$ of the wetting angles through
\begin{equation}
  \label{eq:Q}
  Q = \int_{\theta_a}^{\theta_{P_t}^+} \langle q\rangle \rho(\theta) \diff \theta
    + \int_{\theta_{P_t}^-}^{\theta_b} \langle q\rangle \rho(\theta) \diff \theta
  .
\end{equation}
Inserting \cref{eq:avq} into \cref{eq:Q} gives
\begin{equation}
  \label{eq:Q2}
  Q = -
  \frac{\pi\bar r^4 \sign{\Delta P}}{8\mu L}
  \left[
  \int_{\theta_a}^{\theta_{P_t}^+} 
  \sqrt{\Delta P^2 -\gamma^2} \rho(\theta) \diff \theta
  +
  \int_{\theta_{P_t}^-}^{\theta_b} 
  \sqrt{\Delta P^2 -\gamma^2} \rho(\theta) \diff \theta
  \right]
  .
\end{equation}
Case studies with several different forms of $\rho(\theta)$ will be done numerically in \cref{ssub:numerical_results}.
Here, we will solve \cref{eq:Q2} for a general $\rho(\theta)$ and show that the exponent $\beta$
in
\begin{equation}
  \label{eq:QgoesAs}
  Q \propto \left(\abs{\Delta P} - \abs{P_t}\right) ^ \beta 
\end{equation}
is
\begin{equation}
  \label{eq:beta}
  \beta = 
  \begin{cases}
    1, &\text{ for }
    \abs{\Delta P} - \abs{P_t}
    \gg \abs{\gamma_{\text{max}}} 
    \qquad\qquad\,
    (\text{\cref{cs:beta1}}),\\
    2, &\text{ for }
    \abs{P_t}
    \ll\abs{\Delta P} - \abs{P_t}
    \ll\abs{\gamma_{\text{max}}}
    \quad
    (\text{\cref{cs:beta2}}),\\
    1.5, &\text{ for }
    \abs{\Delta P} - \abs{P_t}
    \ll\abs{P_t}
    \qquad\qquad\quad\:
    (\text{\cref{cs:beta1pt5}}).
  \end{cases}
\end{equation}
Here, $\abs{\gamma_{\text{max}}}$ is the maximum possible threshold pressure a tube can have.

\case{:~~~$\abs{\Delta P} - \abs{P_t}\gg \abs{\gamma_{\text{max}}}$}
\label{cs:beta1}

Using that $\abs{\Delta P}$ is large in this case, we can write 
\begin{equation}
  \label{eq:largeDpLinear}
  \sqrt{\Delta P^2 -\gamma^2} 
  \approx
  \abs{\Delta P}. 
\end{equation}
Inserting \cref{eq:largeDpLinear} into \cref{eq:Q2} and using that the distribution $\rho(\theta)$ is normalized to $1$, gives
\begin{equation}
  \label{eq:Qlinear}
  Q 
  \approx -\frac{\pi\bar r^4 \sign{\Delta P}}{8\mu L}
  \abs{\Delta P}
  \approx -\frac{\pi\bar r^4 \sign{\Delta P}}{8\mu L} \left(\abs{\Delta P} - \abs{P_t}\right)
  .
\end{equation}
In the last step, we have used that $\abs{P_t}$ which is the minimum $\abs{\Delta P}$ needed to achieve $Q>0$, is a much smaller number than $\abs{\Delta P}$.
For the equations derived for all three cases in \cref{eq:beta}, we wish to express $Q$ in terms of $\abs{\Delta P}-\abs{P_t}$ for ease of comparison with \cref{eq:QgoesAs}.
Comparing \cref{eq:Qlinear} with \cref{eq:QgoesAs}, one gets $\beta =1$ for \cref{cs:beta1}.

\vspace{6pt}\noindent\textit{Common for Case 2 and Case 3:}

\Cref{eq:beta} states that the effective pressure obeys $\abs{\Delta P} - \abs{P_t}  \ll\abs{\gamma_{\text{max}}}$ in case~2, while it obeys $\abs{\Delta P} - \abs{P_t} \ll\abs{P_t}$ in case~3.
From \cref{eq:gamma}, the threshold pressure $\gamma$ is so that $\abs{\gamma}\leq k_\gamma$.
This criterion should also be followed by the maximum possible $\abs{\gamma}$ which is $\abs{\gamma_{\text{max}}}$ and the minimum possible $\abs{\gamma}$ which is $\abs{P_t}$. 
Combining these information, a common requirement for cases~2~and~3 should be
\begin{equation}
  \label{eq:requirementCase2n3}
  \abs{\Delta P} - \abs{P_t} \ll k_\gamma.
\end{equation}
\Cref{eq:thetaA} then becomes
\begin{align}
  \nonumber
  \theta_a &= \arccos \left( \frac{\abs {\Delta P}}{k_\gamma} \right),\\
  \nonumber
           &= \arccos \left( \frac{\abs {\Delta P} - \abs{P_t} + \abs{P_t}}{k_\gamma} \right),\\
  \label{eq:thetaA2}
           &\approx\arccos \left( \frac{\abs {P_t}}{k_\gamma} \right)
  - \frac{\abs{\Delta P} - \abs{P_t}}{\sqrt{k_\gamma^2- {P_t^2}}} 
  .
\end{align}
where \cref{eq:requirementCase2n3} was used in the last step.

Next, based on \cref{eq:requirementCase2n3}, Taylor expanding the integrands of \cref{eq:Q2} with respect to $\theta$ around $\theta^\pm_{P_t}$ gives
\begin{align}
  \nonumber
  Q \approx 
  &-\frac{\pi\bar r^4 \sign{\Delta P}}{8\mu L}
  \left[
  \int_{\theta_a}^{\theta_{P_t}^+} 
  \sqrt{\Delta P^2 -P_t^2}\ \rho\left(\theta_{P_t}^+\right) \diff \theta
  \right]
  \\
  &- \frac{\pi\bar r^4 \sign{\Delta P}}{8\mu L}
  \left[
  \int_{\theta_{P_t}^-}^{\theta_b} 
  \sqrt{\Delta P^2 -P_t^2}\ \rho\left(\theta_{P_t}^-\right) \diff \theta
  \right]
  .
\end{align}
Performing this integrations using the integration limits given in \cref{eq:thetaA2,eq:thetaGamma} and that $\theta_b = 180^\circ - \theta_a$ (\cref{eq:thetaB}) gives
\begin{align}
  \nonumber
   Q \approx - 
   &\left[
  \frac{\pi\bar r^4 \sign{\Delta P}}{8\mu L}
  \frac{\rho\left(\theta_{P_t}^+\right) + \rho\left(\theta_{P_t}^-\right)}{\sqrt{k_\gamma^2 - P_t^2  }} 
  \right]
  \\
  \label{eq:QsmallP}
  &\times\left(\abs{\Delta P} - \abs{P_t}\right) ^ {3/2}
  \left(\left(\abs{\Delta P} - \abs{P_t}\right) + 2 \abs{P_t}\right) ^ {1/2}
  .
\end{align}
\Cref{eq:QsmallP} holds true for both \cref{cs:beta2,cs:beta1pt5}.

\case{:~~~$\abs{P_t} \ll\abs{\Delta P} - \abs{P_t}  \ll\abs{\gamma_{\text{max}}}$}
\label{cs:beta2}

The part of the criterion for this case that says $\abs{P_t}\ll\abs{\Delta P} - \abs{P_t}$ makes it so that $\left(\left(\abs{\Delta P} - \abs{P_t}\right) + 2 \abs{P_t}\right) ^ {1/2} \approx \left(\abs{\Delta P} - \abs{P_t}\right) ^ {1/2}$ which can be used to write \cref{eq:QsmallP} as
\begin{equation}
  \label{eq:Qbeta2}
   Q \approx -
   \left[
  \frac{\pi\bar r^4 \sign{\Delta P}}{8\mu L}
  \frac{\rho\left(\theta_{P_t}^+\right) + \rho\left(\theta_{P_t}^-\right)}{\sqrt{k_\gamma^2 - P_t^2  }} 
  \right]
  \left(\abs{\Delta P} - \abs{P_t}\right)^2 
  .
\end{equation}
Comparing \cref{eq:Qbeta2} to \cref{eq:QgoesAs} gives $\beta =2$ for \cref{cs:beta2}.

\case{:~~~$\abs{\Delta P} - \abs{P_t} \ll\abs{P_t}$}
\label{cs:beta1pt5}

The criterion for this case makes it so that $\left(\left(\abs{\Delta P} - \abs{P_t}\right) + 2 \abs{P_t}\right) ^ {1/2} \approx \sqrt{2 \abs{P_t}}$ which can be used to write \cref{eq:QsmallP} as
\begin{equation}
 \label{eq:Qbeta1pt5}
   Q \approx 
  - \left[
  \frac{\pi\bar r^4 \sign{\Delta P}}{8\mu L}
  \frac{\rho\left(\theta_{P_t}^+\right) + \rho\left(\theta_{P_t}^-\right)}{\sqrt{k_\gamma^2 - P_t^2  }} 
  \sqrt{2 \abs{P_t}}
  \right]
  \left(\abs{\Delta P} - \abs{P_t}\right) ^ {3/2}
  .
\end{equation}
Comparing \cref{eq:Qbeta1pt5} to \cref{eq:QgoesAs} gives $\beta = 3/2$ for \cref{cs:beta1pt5}.
\newline

\begin{figure}[ht]
  \centering
  \includegraphics[width=0.5\linewidth]{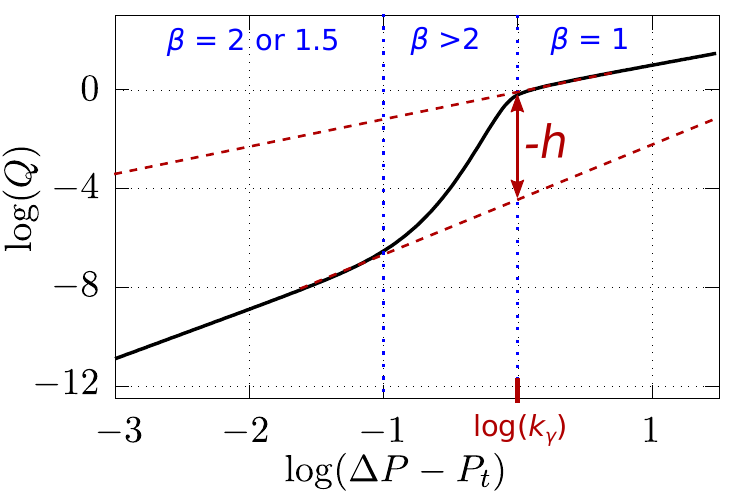}
  \caption{Height $h$ of the transition region is defined as the distance between the two lines that are extrapolations of the linear parts of the curve at $\Delta P-P_t=k_\gamma$. Here, $Q$ is the volumetric flow rate and $\Delta P-P_t$ is the excess pressure.}
  \label{fig:CFB_trans}
\end{figure}

In the transition region between \cref{cs:beta1} and \cref{cs:beta1pt5,cs:beta2}, where $\abs{\Delta P} - \abs{P_t} \approx k_\gamma$, the volumetric flow depends more sensitively on the wetting angle distribution function $\rho(\theta)$, and is in general not a simple power law.
Nevertheless, we can use the analysis presented here to compute the height of that transition region. 
Taking the logarithm of $Q$ we find that
\begin{equation}
  \log(Q) = \begin{cases}
    \eta + \log\left(\frac{\abs{\Delta P}}{k_\gamma} - \frac{\abs{\gamma}}{k_\gamma}\right)  & \text{for \cref{cs:beta1}}, \\
    \eta + 2\log\left(\frac{\abs{\Delta P}}{k_\gamma} - \frac{\abs{\gamma}}{k_\gamma}\right) + \log\left(\frac{\rho\left(\theta_{P_t}^+\right) + \rho\left(\theta_{P_t}^-\right)}{\sqrt{1 - (P_t/k_\gamma)^2}}\right) & \text{for \cref{cs:beta2}}, \\
    \begin{aligned}[b]
    \eta + 1.5\log\left(\frac{\abs{\Delta P}}{k_\gamma} - \frac{\abs{\gamma}}{k_\gamma}\right) + \log\left(\frac{\rho\left(\theta_{P_t}^+\right) + \rho\left(\theta_{P_t}^-\right)}{\sqrt{1 - (P_t/k_\gamma)^2}}\right) \\
    + \log\left(\sqrt{\frac{2\abs{P_t}}{k_\gamma}}\right)
    \end{aligned} & \text{for \cref{cs:beta1pt5}},
  \end{cases}
\end{equation}
where $\eta$ is a constant that is the same for all the three cases.
Evaluating this at $\abs{\Delta P} - \abs{P_t} = k_\gamma$ we see that the height difference between \cref{cs:beta2} and \cref{cs:beta1} in a logarithmically scaled plot is
\begin{equation}
  h = \log\left(\frac{\rho\left(\theta_{P_t}^+\right) + \rho\left(\theta_{P_t}^-\right)}{\sqrt{1 - (P_t/k_\gamma)^2}}\right),
  \label{eq:height_21}
\end{equation}
and the height difference between \cref{cs:beta1pt5} and \cref{cs:beta1} is
\begin{equation}
  h = \log\left(\frac{\rho\left(\theta_{P_t}^+\right) + \rho\left(\theta_{P_t}^-\right)}{\sqrt{1 - (P_t/k_\gamma)^2}}\right)
    + \log\left(\sqrt{\frac{2\abs{P_t}}{k_\gamma}}\right).
\end{equation}
This is shown in \cref{fig:CFB_trans}.
It is assumed in the above analysis that $\rho\left(\theta_{P_t}^+\right) + \rho\left(\theta_{P_t}^-\right) < \sqrt{1 - (P_t/k_\gamma)^2}$, since $\rho(\theta)$ is a normalized probability distribution.
Hence, $h<0$, which means that the $\beta$ in the transition region must be larger than outside the transition region.

\subsubsection{Numerical Results}%
\label{ssub:numerical_results}

Here, we study numerically the volumetric flow rate $Q$'s response to a wide range of an applied pressures $\Delta P$.
In addition to verify the analytical results from \cref{ssub:a_bundle_of_tubes}, this numerical study allows us to probe the transition region where $\Delta P$ is the same order of magnitude as $\gamma_\text{max}$.
We present results with three different normal and uniform distributions of $\theta$, and note that we find similar results also for many other distributions.

\begin{figure}[ht]
\centering
\subcaptionbox{\label{fig:CFB_exUni}}{\includegraphics[width=0.50\textwidth]{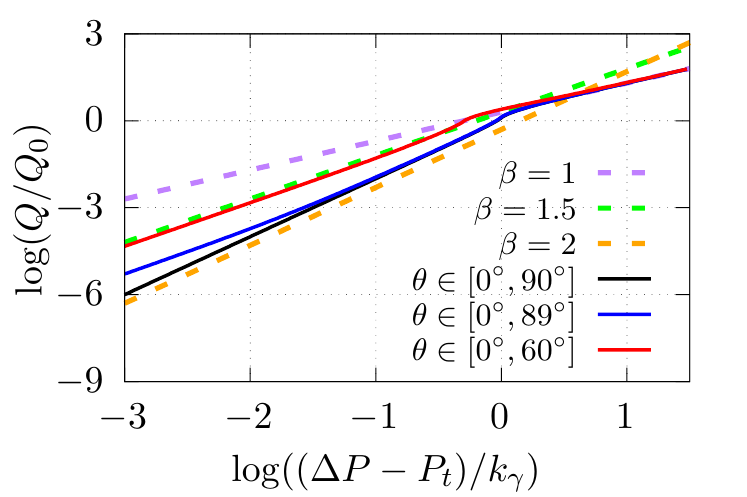}}%
\hfill
\subcaptionbox{\label{fig:CFB_exABC}}{\includegraphics[width=0.50\textwidth]{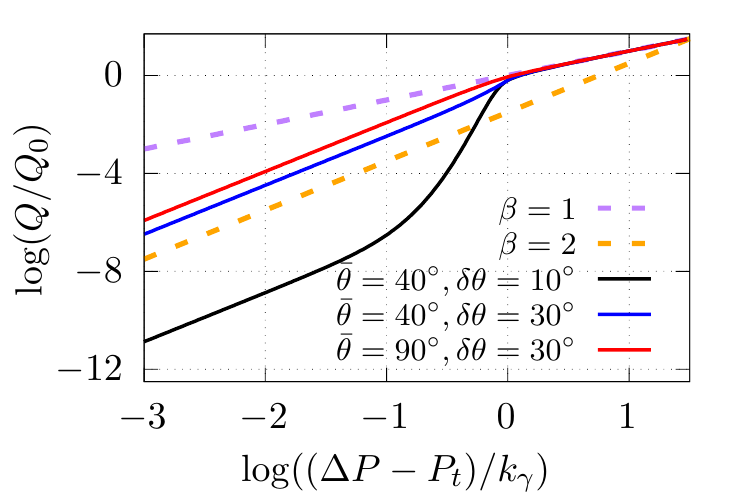}}%
\hfill
\caption{Relation between the total volumetric flow rate $Q$ and the excess pressure $\Delta P - P_t$ for wetting angles $\theta$ (a) uniformly distributed between $[0^\circ,90^\circ]$, $[0^\circ,89^\circ]$ and $[0^\circ,60^\circ]$, and (b) normally distributed with $(\bar\theta=40^\circ, \delta\theta=40^\circ)$, $(\bar\theta=40^\circ, \delta\theta=40^\circ)$ and $(\bar\theta=40^\circ, \delta\theta=40^\circ)$ where $\bar\theta$ is the mean and $\delta\theta$ is the standard deviation. 
Here, $\Delta P$ is the applied pressure and $P_t$ is the minimum threshold pressure. 
The flow rates have been normalized with $Q_0=-(\pi\bar r^4 \sign{\Delta P}k_\gamma)/(8\mu L)$ and the pressures have been normalized with $k_\gamma$.
Straight lines with gradients $\beta$ has been added.
}%
\label{fig:CFB}
\end{figure}

First, consider $\rho(\theta)$ to be a uniform distribution,
\begin{equation}
  \label{eq:uniformRho}
  \rho(\theta) = 
  \begin{cases}
    \frac{1}{\theta_b-\theta_a} , &\text{ for }\theta\in[\theta_a,\theta_b]
    \text{ with } 0^\circ <(\theta_a,\theta_b)<180^\circ, \\
    0, &\text{ otherwise }.
  \end{cases}
\end{equation}
The results from when $\theta$ is uniformly distributed between $[\theta_a=0^\circ,\theta_b=90^\circ]$, $[\theta_a=0^\circ,\theta_b=89^\circ]$ and $[\theta_a=0^\circ,\theta_b=60^\circ]$ are shown in \cref{fig:CFB_exUni}.
These results confirm the analytical calculations performed in \cref{ssub:a_bundle_of_tubes}.
When $\abs{\Delta P} - \abs{P_t}\gg \abs{\gamma_{\text{max}}}$, \cref{cs:beta1}, all three examples do indeed satisfy $\beta=1$, reflecting linear Darcy flow.
The region where $\abs{P_t}\ll\abs{\Delta P} - \abs{P_t}\ll\abs{\gamma_{\text{max}}}$ resulting in $\beta = 2$, \cref{cs:beta2}, covers the rest of the plot for $\rho(\theta)$ with $\theta \in[0^\circ,90^\circ]$, since in this case $P_t = 0$.
For $\rho(\theta)$ with $\theta \in[0^\circ,89^\circ]$ however, that same region is approximately only the center part of the plot $(-1<\log((\Delta P-P_t)/k_\gamma)<0)$ while the rest belongs to the regime where $\beta =1.5$.
The region $\abs{\Delta P} - \abs{P_t}\ll\abs{P_t}$, \cref{cs:beta1pt5}, that gives $\beta =1.5$ is dominating in the case with $\theta \in[0^\circ,60^\circ]$.
For this distribution, the transition from $\beta =1$ to $\beta =1.5$ happens quickly due to $\abs{P_t} = k_\gamma\cos(60^\circ)$ being a larger number than in the other two cases, rendering the region where the applied pressure can satisfy the requirement for $\beta=2$, namely $\abs{P_t}\ll\abs{\Delta P} - \abs{P_t}\ll\abs{\gamma_{\text{max}}}$, very small.

The CFB model has also been numerically tested with $\rho(\theta)$ being normal distribution with mean $\bar\theta$ and standard deviation $\delta\theta$ given by 
\begin{equation}
  \label{eq:normalRho}
  \rho(\theta) = 
    \frac{\exp\left\{-\frac{1}{2} \left( \frac{\theta -\bar\theta}{\delta\theta} \right)^2 \right\}}{\int_{0^\circ}^{180^\circ}\exp\left\{-\frac{1}{2} \left( \frac{\theta -\bar\theta}{\delta\theta} \right)^2 \right\}\diff \theta } 
    \ 
    \quad\text{ for }\theta\in[0^\circ,180^\circ] \\
.
\end{equation}
The results from when $\theta$ is normally distributed with $(\bar\theta=40^\circ, \delta\theta=10^\circ)$, $(\bar\theta=40^\circ, \delta\theta=30^\circ)$ and $(\bar\theta=90^\circ, \delta\theta=30^\circ)$ are shown in \cref{fig:CFB_exABC}.
These results once again confirm the analytical calculations performed in \cref{ssub:a_bundle_of_tubes}.
Toward the right in \cref{fig:CFB_exABC} where $\abs{\Delta P} - \abs{P_t}\gg \abs{\gamma_{\text{max}}}$, \cref{cs:beta1}, all three examples follow $\beta=1$.
Toward the left in \cref{fig:CFB_exABC} where $\abs{P_t}\ll\abs{\Delta P} - \abs{P_t}\ll\abs{\gamma_{\text{max}}}$, the results follow $\beta = 2$, \cref{cs:beta2}.
Notice that, $P_t=0$ for all these three cases. 
Since a nonzero $Q$ occurs only when $\abs{\Delta P} > \abs{P_t}$, the requirement for $\beta =1.5$, namely $\abs{\Delta P} - \abs{P_t}\ll\abs{P_t}$, \cref{cs:beta1pt5}, is not satisfied here. 
In \cref{fig:CFB_exABC}, the transition region between when $\beta=1$ and when $\beta=2$ exhibits a gradient $\beta>2$, in accordance with the analysis above.
The same effect can also slightly be seen in \cref{fig:CFB_exUni}, but is much more apparent in \cref{fig:CFB_exABC}. 
One can also see from \cref{fig:CFB_exABC} that $\beta$ is larger for smaller $\delta\theta$ and means $\bar\theta$ further away from $90^\circ$.
This can be understood from \cref{eq:height_21}, since smaller $\delta\theta$ and larger $\abs{\bar\theta - 90^\circ}$ implies smaller $\rho(90^\circ)$.
From \cref{eq:height_21}, we see that smaller $\rho(90^\circ)$ implies a larger height difference $h$.
Physically, this can be understood from the fact that a smaller $\rho(90^\circ)$ means that a smaller fraction of the total number of tubes are active in the low pressure regime.

\subsubsection{The Origin of $\beta$}%
\label{ssub:the_origin_of_beta}

We can write volumetric flow rate as $Q = N \bar{ q}$, where $N$ is the number of open tubes and $\bar{q}$ is the average flow per open tube.
Put differently, $\bar{q}$ is the average of $\langle q\rangle$ for all the open tubes.
We propose that $\beta$ can be understood from how each of these factors change with applied pressure $\Delta P$.
Suppose that increasing the pressure difference from $\abs{\Delta P_0}-\abs{P_t}$ to $\abs{\Delta P}-\abs{P_t} = \bar x\left(\abs{\Delta P_0}-\abs{P_t}\right)$ transforms the number of open tubes and the flow per tube according to $N_0 \to \bar x^{\bar{a}} N_0$ and $\bar q_0 \to \bar x^{\bar{b}} \bar q_0$, respectively.
In this case we see that the volumetric flow change from $Q_0 = N_0\bar q_0$ to 
\begin{equation}
  Q = \left(\frac{\abs{\Delta P}-\abs{P_t}}{\abs{\Delta P_0}-\abs{P_t}}\right)^{\bar{a}+\bar{b}}Q_0,
\end{equation}
so $\beta = \bar{a} + \bar{b}$.

Consider first $\bar q$.
From \cref{eq:avq}, we know that the volumetric flow for a tube with threshold pressure $\gamma$ is $\langle q\rangle \propto \sqrt{\Delta P^2 -\gamma^2} = \sqrt{(\Delta P - \gamma)(\Delta P - \gamma + 2\gamma)}$.
Thus, we see that $\langle q \rangle \propto (\Delta P - \gamma)^{1/2}$ if $(\Delta P - \gamma) \ll \gamma$ and $\langle q \rangle \propto (\Delta P - \gamma)^1$ if $(\Delta P - \gamma) \gg \gamma$.
Hence, if most of the active tubes have threshold pressure just below the applied pressure (\cref{cs:beta1pt5}), then $\bar{a} = 1/2$.
On the other hand, if most of the active tubes have threshold pressure well below the applied pressure (\cref{cs:beta1,cs:beta2}), then $\bar{a} = 1$.

Next, consider the number of active tubes transporting fluid, N.
This is given by
\begin{equation}
  \label{eq:Ndp}
  N = N_\text{max} 
  \left[
  \int_{\theta_a}^{\theta_{P_t}^+} 
  \rho(\theta) \diff \theta
  +
  \int_{\theta_{P_t}^-}^{\theta_b} 
  \rho(\theta) \diff \theta
  \right]
  , 
\end{equation}
where $N_\text{max}$ is the total number of tubes in the system.
When the applied pressure is larger than the maximal threshold pressure $\gamma_\text{max}$, then all the tubes are active and $N = N_\text{max}$.
Thus, for \cref{cs:beta1}, we have $\bar{b} = 0$ and consequently $\beta = \bar{a} + \bar{b} = 1$.
On the other hand, when $\abs{\Delta P} - \abs{P_t} \ll \abs{\gamma_\text{max}}$, then $\theta_{P_t}^+ - \theta_a = \theta_b - \theta_{P_t}^- = \left(\abs{\Delta P} - \abs{P_t}\right)/\sqrt{k_\gamma^2 - P_t^2} \ll 1$, as seen from \cref{eq:thetaA2}.
Thus,
\begin{equation}
  N = \frac{N_\text{max}(\rho(\theta_a)+\rho(\theta_b))}{\sqrt{k_\gamma^2 - P_t^2}}\left(\abs{\Delta P} - \abs{P_t}\right),
\end{equation}
so $\bar{b} = 1$.
Combining this with the result for $\bar{a}$ we see that in \cref{cs:beta2}, we get $\beta = \bar{a}+\bar{b}=2$, and in \cref{cs:beta1pt5}, we get $\beta = \bar{a}+\bar{b}=1.5$.
This explains why $\beta \in \{1, 1.5, 2\}$ when either all tubes are active (\cref{cs:beta1}) or only a small fraction is active (\cref{cs:beta2,cs:beta1pt5}).

\subsection{The Dynamic Pore Network Model Results}%
\label{sub:the_dynamic_pore_network_model_results}

\subsubsection{Data Collecting and Processing Procedures}%
\label{ssub:data_collecting_and_processing_procedures}

Using the method described in \cref{sub:the_dynamic_pore_network_model_description}, numerical simulations of the DPN model have been performed.
The following factors and parameters have been kept constant during all simulations.
The 2D network used was made of $64\times 64$ links and had periodic boundary conditions in all directions.
All the links had length $l=1$~mm, average radii $\bar r\in[0.1l,0.4l]$ and amplitude of the periodic variation $a=1$~mm, see \cref{eq:radius}.
The viscosities of the fluids $A$ and $B$ were $\mu_A=\mu_B=0.01$~Pa$\cdot$s, see \cref{eq:mu}.
The surface tension between the fluids were $0.03$~N/m, see \cref{eq:pc}.
In this closed network system, the control parameter was the saturation of one of the fluids in the whole system, which was tested for values $S_A\in\{0.1,0.2,0.3,0.4,0.5,0.6,0.7,0.8,0.9\}$.

The distributions of the wetting angles, $\rho(\theta)$, have firstly been tested for a uniform distribution with $\theta\in [0^\circ,180^\circ]$.
Secondly, normal distributions with means $\bar\theta\in\{0^\circ,30^\circ,60^\circ,90^\circ\}$ and, with each mean, standard deviations $\delta\theta\in\{0^\circ,30^\circ,60^\circ\}$ were also tested.
The $\rho(\theta)$ mentioned until now were implemented into the network with all of the above-mentioned values of $S_A$.
In addition to this, at $S_A=0.5$, several more normally distributed $\theta$ were examined.
They were with means $\bar\theta\in\{0^\circ,30^\circ,60^\circ,90^\circ\}$ and, with each mean, standard deviations $\delta\theta\in\{15^\circ,45^\circ,75^\circ\}$, as well as, with means $\bar\theta\in\{15^\circ,45^\circ,75^\circ\}$ and, with each mean, standard deviations $\delta\theta\in\{0^\circ,30^\circ,60^\circ\}$.
Note that the normally distributed $\theta$ could go outside the interval $[0^\circ,180^\circ]$, which is equivalent to a slightly increased weight around $0^\circ$ or $180^\circ$ because the angle only comes in through $\cos\theta$, as seen in \cref{eq:pc}.
Another thing to note is that only distributions with $\bar\theta \leq 90^\circ$ have been considered.
This is because since the fluids have the same viscosity, a symmetry is in place where the case with mean $\bar\theta$ and saturation $S_A$ is the same as the case with mean $180^\circ -\bar\theta$ and saturation $1-S_A$.

For each $S_A$, the system was driven by various different $Q$, and for each $Q$, $20$ different realizations of the network were performed and averaged over.
The global applied pressure $\Delta P$ in the direction of the flow was measured, and was calculated by averaging over the fluctuations after the system had reached a steady state.
After obtaining a set of $Q$ and $\Delta P$ for every $\rho(\theta)$ at every $S_A$, data analysis had to be performed to determine the global threshold pressure $P_t$ below which there is no flow through the whole system, as well as the exponent $\beta$ in $Q \propto \left(|\Delta P|-|P_t|\right)^{\beta}$.
Note that similar to the CFB model, the first active path across the entire system in DPN occurs once $\Delta P$ exceeds $P_t$.

\begin{figure}[ht]
  \centering
  \includegraphics[width=0.7\linewidth]{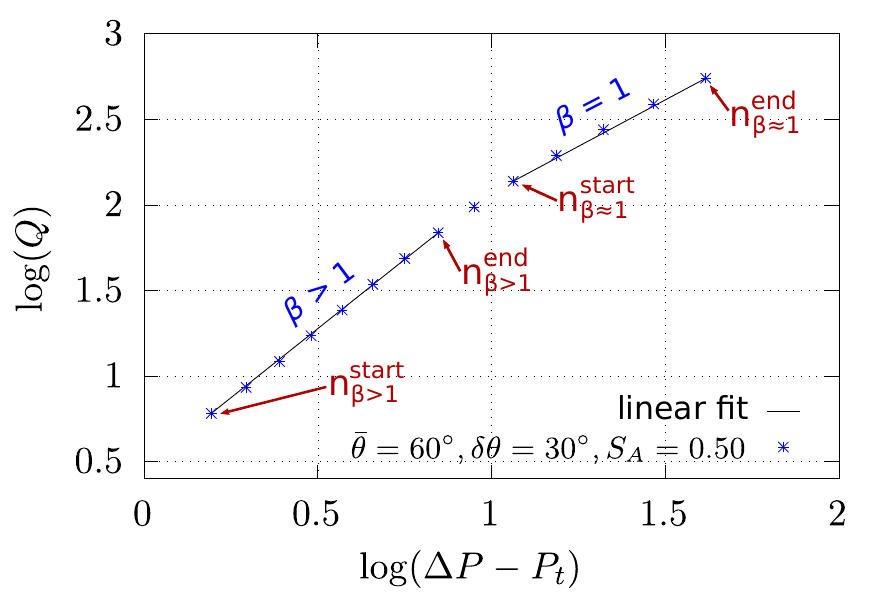}
  \caption{ Linear fitting the data points from index $n_{\beta\approx 1}^\text{start}$ to $n_{\beta\approx 1}^\text{end}$ and data points from index $n_{\beta>1}^\text{start}$ to $n_{\beta>1}^\text{end}$ separately gives $\beta$ in those regions.}
  \label{fig:DPN_exInk}
\end{figure}

The process of determining $P_t$ and $\beta$ started with deciding the indices of the data points that belonged to the linear and power law regimes through visual examination.
As shown with an example in \cref{fig:DPN_exInk}, this meant deciding the indices of the datapoints that lied between $n_{\beta\approx 1}^\text{start}$ to $n_{\beta\approx 1}^\text{end}$ and belonged to the region with $\beta\approx 1$, as well as the indices between $n_{\beta>1}^\text{start}$ to $n_{\beta>1}^\text{end}$ that belonged to the region with $\beta>1$.
The error bars were calculated as the absolute values of the difference between the results and the results that would have been if the range of data points included from each region were reduced.
The next step was to perform linear fitting of $Q^{1/\beta}$ against $\Delta P$ on the data points from $n_{\beta>1}^\text{start}$ to $n_{\beta>1}^\text{end}$. 
The linear fitting was of the form $c_1 Q^{1/\beta} + c_2$ with $c_1$ and $c_2$ real numbers.
Due to the definition that $P_t = \Delta P$ exactly when $Q^{1/\beta}$ becomes nonzero, $P_t=c_2$.
This procedure was repeated for a range of different $\beta$s, and the $P_t$ that gave the least root-mean-square error was chosen as the final candidate.
Thereafter, linear fitting $\log(Q)$ versus $\log(\Delta P - P_t)$ separately for the data points from index $n_{\beta\approx 1}^\text{start}$ to $n_{\beta\approx 1}^\text{end}$ and data points from index $n_{\beta>1}^\text{start}$ to $n_{\beta>1}^\text{end}$ gave the values of $\beta$ in those regions.

\subsubsection{Simulation Results}%
\label{ssub:simulation_results}

All of the simulations performed, using the parameters and the different wetting angle distributions $\rho(\theta)$ described in \cref{ssub:data_collecting_and_processing_procedures}, resulted in a Darcy-like flow with $\beta \approx 1$ in the high pressure limit where most of the links were active. 
The transition region with $\beta>2$ flow, as in the case with the CFB model, was not observed with the DPN model.
Note that the DPN model, compared to the CFB model, simulates a porous medium that has a more complex interplay of the fluids in the links which separate and rejoin. 
It could be speculated that this advanced behavior of the network eliminates the transition region originally observed in the CFB model, in other words, the transition region may be an artifact of the CFB model. 
In the low pressure limit result of the DPN model, the exponent $\beta$ shows dependence on the saturation and the wettability properties of the network.
This is also the case with the threshold pressure of the network $P_t$.
A closer exploration of the latter two factors will now be presented.

\begin{figure}[ht]
  \centering
  \includegraphics[width=1.0\linewidth]{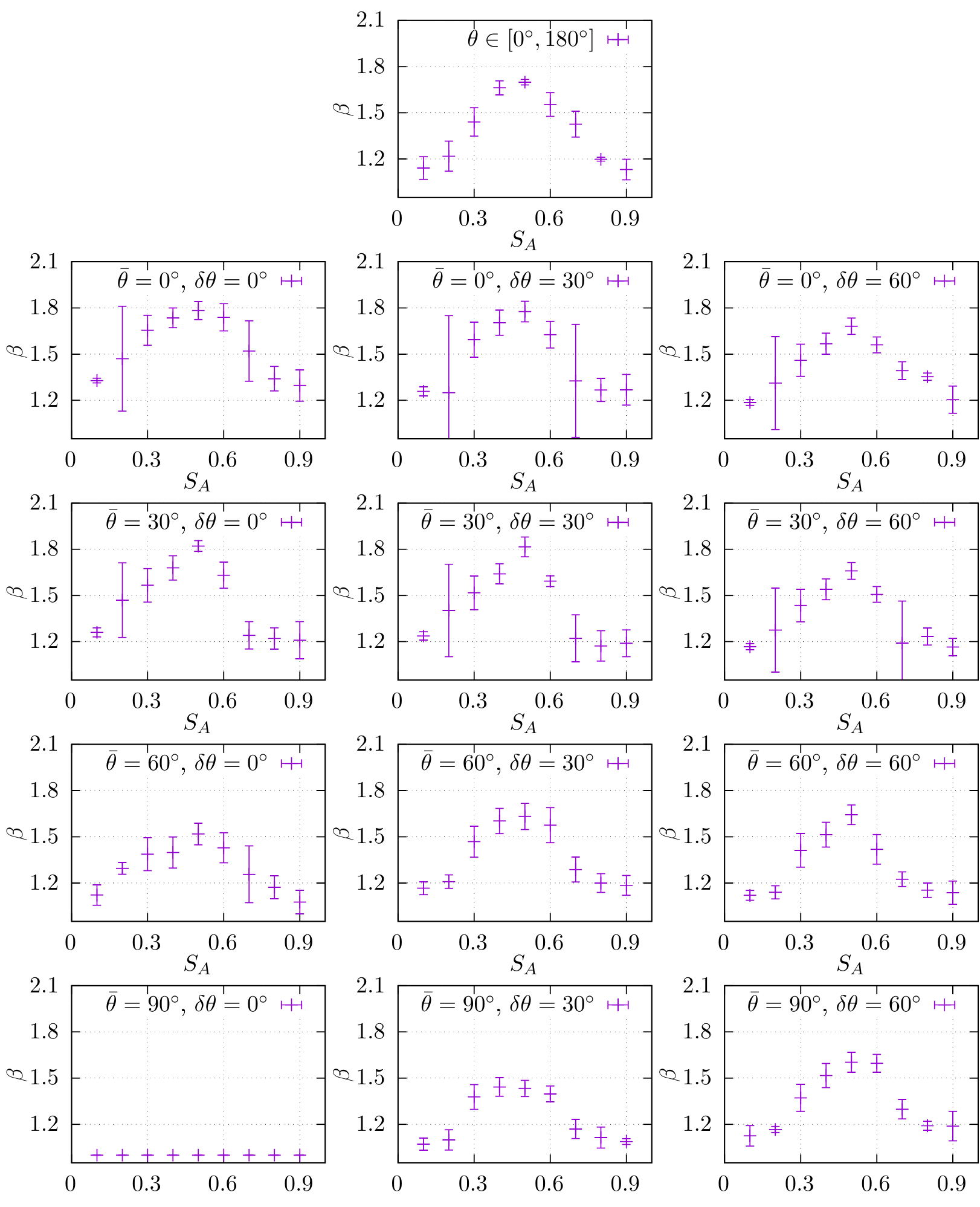}
  \caption{Dependence of $\beta$ in the low pressure limit on the saturation of one of the fluids $S_{A}$ when the distributions of the wetting angles $\rho(\theta)$ is uniform distribution with $\theta\in[0^\circ,180^\circ]$ (uppermost) and normal distributions with means $\bar\theta\in\{0^\circ,30^\circ,60^\circ,90^\circ\}$ and standard deviations $\delta\theta\in\{0^\circ,30^\circ,60^\circ\}$.}
  \label{fig:SvsBeta}
\end{figure}

The results for $\beta$ in the \textit{low pressure limit} are shown in \cref{fig:SvsBeta} and takes on various values in the range $\beta\in[1.00\pm 0.05,1.82\pm 0.05]$.
The phenomenon with $\beta > 1$ originates from that many links in the network are not yet opened in the low pressure regime, which means increasing $\Delta P$ increases the number of active links in addition to increasing the flow within each active link. 
The overall combined effect of these allows the volume of fluid transported to rise much more than if all the links were already open.
This is the same as in the capillary fiber bundle model, but here $\beta$ takes on a larger range of values depending on the saturation and the wetting conditions.

\begin{figure}[ht]
  \centering
  \includegraphics[width=1.0\linewidth]{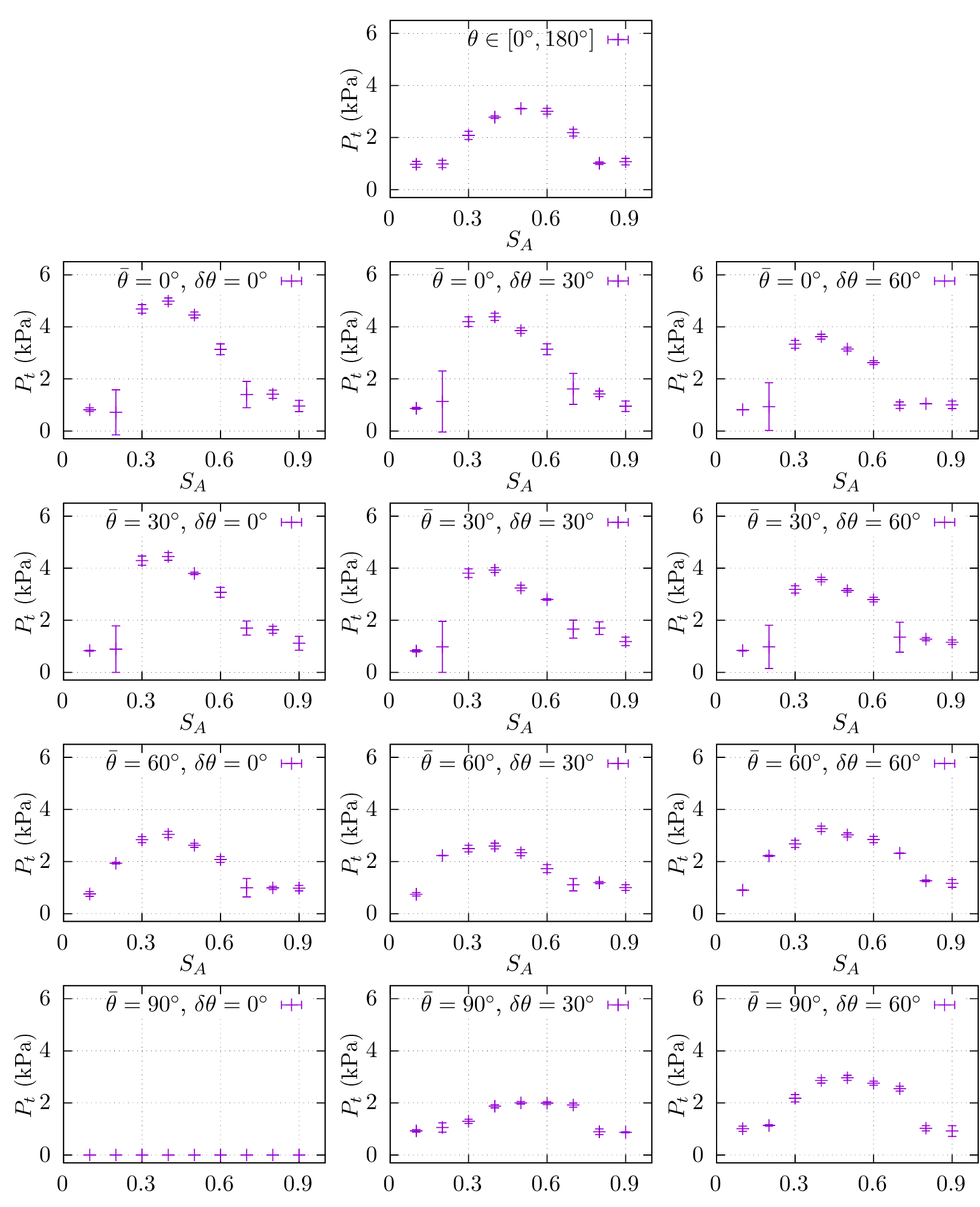}
  \caption{Dependence of the threshold pressures $P_t$ on the saturation of one of the fluids $S_{A}$ when the distributions of the wetting angles $\rho(\theta)$ is uniform distribution with $\theta\in[0^\circ,180^\circ]$ (uppermost) and normal distributions with means $\bar\theta\in\{0^\circ,30^\circ,60^\circ,90^\circ\}$ and standard deviations $\delta\theta\in\{0^\circ,30^\circ,60^\circ\}$.}
  \label{fig:SvsPt}
\end{figure}

The results for $P_t$ are shown in \cref{fig:SvsPt}.
The exponents $\beta$ in \cref{fig:SvsBeta}, as well as, the minimum threshold pressures $P_t$ in \cref{fig:SvsPt} have a tendency to be largest for saturations around $0.5$ and decrease steadily with increasing saturation of either one of the fluids.
The reason is that when one of the fluids dominates the system, $S_A \rightarrow 0.1$ or $0.9$, it is easy for that dominating fluid to create an active flow-path through the system.
This is because those connected links that contain the same fluid will not experience a interfacial capillary pressure barrier.
This decreases the overall threshold $P_t$ of the system, which is the cumulative effect of the interfacial capillary barriers in the network.
There will be few new links to become active as $\Delta P$ increases under these circumstances which will further make $Q$ less reactive toward changes in $\Delta P$, meaning a decreased $\beta$. 
In contrast, when there are comparable amounts of the fluids $A$ and $B$ in the system, $S_A \rightarrow 0.5$, $\Delta P$ has to overcome the cumulative capillary pressure barrier created by the large number of interfaces between $A$ and $B$.
This naturally has an increasing effect on $P_t$. 
When $\Delta P$ is increased under such conditions, the requirement for nonzero flow for many links are satisfied at once, causing a drastic increase in $Q$ as a response, which increases $\beta$.
Lastly regarding the effect of saturation, in the three middle rows in \cref{fig:SvsPt}, the maxima of the $P_t$ plots are skewed to the left of $S_A=0.5$.
In those cases, $\theta$ is concentrated around a $\bar\theta<90^\circ$, which makes fluid $A$ is the most wetting fluid while $B$ is the most non-wetting fluid.
It is easier for the wetting fluid to get transported in a porous medium, meaning when the saturation of the wetting fluid is lower, $S_A<0.5$, the system will require a higher applied pressure to achieve a non-zero flow making the system's $P_t$ higher.

\begin{figure}[ht]
\centering
\subcaptionbox{\label{fig:mu_beta5}}{\includegraphics[width=0.50\textwidth]{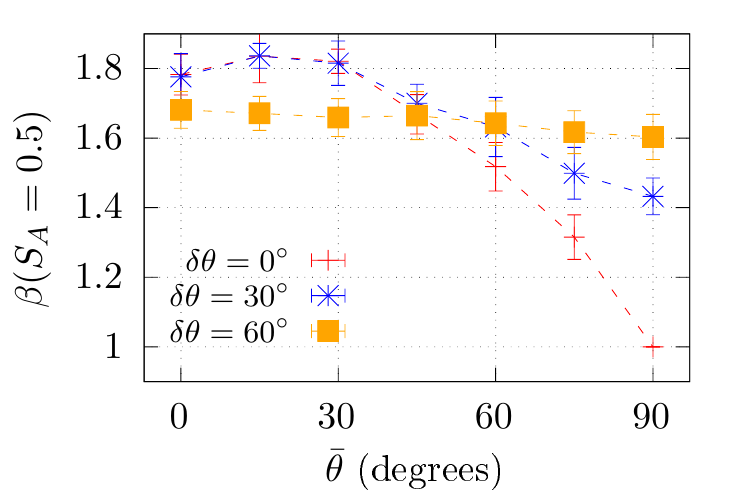}}%
\hfill
\subcaptionbox{\label{fig:sigma_beta5}}{\includegraphics[width=0.50\textwidth]{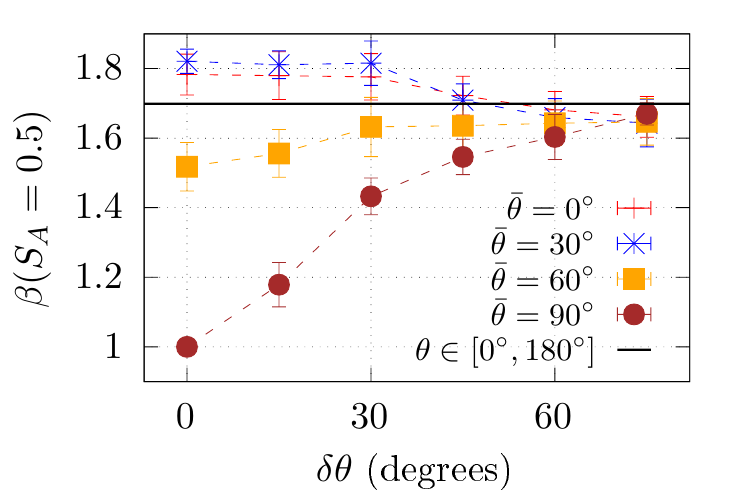}}%
\hfill
\caption{The values of $\beta$ at $S_A=0.5$ as functions of (a) the mean $\bar\theta$ and (b) the standard deviation $\delta\theta$ of the wetting angle distribution. $\theta\in[0^\circ,180^\circ]$ in (b) indicates uniformly distributed wetting angles.
}%
\label{fig:muSigma_beta5}
\end{figure}

The variations in $\beta$ for different $\rho(\theta)$ are more subtle than in the case with $P_t$. 
To get a clear overview of the differences, the maximum values, $\beta(S_A=0.5)$, have been plotted as functions of $\bar\theta$ and $\delta\theta$ in \cref{fig:mu_beta5,fig:sigma_beta5}, respectively.
Both the exponents $\beta$ in \cref{fig:mu_beta5,fig:sigma_beta5} and the minimum threshold pressure $P_t$ in \cref{fig:SvsPt} vary with $\bar\theta$ and $\delta\theta$ of the normal distributions.
The interfacial capillary pressures, given by \cref{eq:pc}, increase with the distance between the wetting angles and $90^\circ$.
This happens with increasing $\delta\theta$ for wetting angles with mean around $\bar\theta \approx 90^\circ$, or with decreasing $\delta\theta$ for $\bar\theta$ deviating from $90^\circ$.
Note that a larger $\delta\theta$ means that the wettability is allowed to deviate more from $\bar\theta$.
This reflects in the values of $P_t$ in \cref{fig:SvsPt} where the peaks of the plots in row~2~and~3 decrease from left to right while the peaks in row~5 increase from left to right, and the peaks decrease from top to bottom.
The same effect also creates the trend of decreasing $\beta$ as $\bar\theta\rightarrow 90^\circ$ in \cref{fig:mu_beta5}.
When links have a wetting angle $\theta$ close to $90^\circ$, many links will open at very small pressure $\Delta P$, which means increasing $\Delta P$ in the typical low pressure limit does not open many new links, hence raises $Q$ with a small $\beta$.
\Cref{fig:sigma_beta5} also supports this phenomena.
\Cref{fig:sigma_beta5}, in addition, shows the expected result that as $\delta\theta$ increases, $\beta$ for the various normal distributions approach a value close to that of the uniform distribution with $\theta\in[0^\circ,180^\circ]$.
Instead of having a small range of wetting angles around a $\bar\theta$, uniform distribution provides wetting angles anywhere between $0^\circ$ to $180^\circ$ with equal probability.
Therefore, it makes sense that uniform distribution results in $\beta$ and $P_t$ values most similar to those of normal distributions with largest $\delta\theta$ which have the most variation in the wetting angles.

From the results presented here we see that both $P_t$ and $\beta$ depend on the wetting angle distribution $\rho(\theta)$.
In particular, they are affected not only by the mean wetting angle, but also by the spread of wetting angles.
Thus, in order to fully characterize the flow through a porous media it is in general insufficient to assume a uniform wettability.


\section{Conclusion}%
\label{sec:conclusion}

We studied systematically the effect of mixed wetting conditions on the effective rheology of two-phase flow in porous media by using the capillary fiber bundle (CFB) model and the dynamic pore network (DPN) model.
Although the two models are not quantitatively comparable, they are qualitatively similar.
Both models show that mixed wettability conditions can have significant influence on the rheology in terms of the dependence of the global volumetric flow rate $Q$ on the global pressure drop $\Delta P$. 
In the CFB model, the effect of mixed wettability, in other words the shape of the wetting angle distribution, plays the most significant role in the transition regime between low and high $\Delta P$ limit where most of the tubes opens.
In the DPN model, the whole process leading up to opening of all the possible links produces a $Q$ that depends on the wettability of the system. 
Hence, the studies carried out in this work show that the behavior of immiscible two-phase flow in porous media changes when we move from uniform to mixed wet conditions.
The wettability distribution of the porous media is therefore an important factor that should be taken into account when studying the rheology in porous media.
Future works may study spatial correlation in the wettability distributions, as well as, the effect of varying viscosities of the fluids, which have not been done in this study.

From the CFB model, we found that the exponent $\beta$ in $Q \propto \left(\abs{\Delta P} - \abs{P_t}\right) ^ \beta$ is $1$, $2$ and $3/2$ in the high, low and very low effective pressure limits, respectively.
The numerical solutions in addition revealed the $\beta>2$ behavior in the transition region between these extreme limits, which was due to the rapid opening of tubes in that pressure range.
In this model, the functional form of the wetting angle distribution $\rho(\theta)$ has the largest effect in the transition region, since this is the region where most tubes become active.

In the DPN model, on the other hand, we found that $\rho(\theta)$ is influential for the behavior at all pressure drops below the Darcy regime.
The transition region with $\beta>2$ flow could not be observed with the DPN model, leading to the speculation that the transition region could have been an artifact of the CFB model.
In the low pressure limit, $\beta$ had values varying anywhere between $\beta\in[1.00\pm 0.05,1.82\pm 0.05]$.
Both $\beta$ in this pressure limit and the threshold pressure $P_t$ showed the tendency to be largest for saturations around $0.5$ and decrease steadily with increasing saturation of either one of the fluids.
The reason is that when there is comparable amount of both fluids in the system, there will be large number of interfaces in the system giving large interfacial capillary pressure for the link. 
This works to increase $P_t$ as well as making $Q$ more reactive toward changes in $\Delta P$, which increases $\beta$.
Finally, we found that $\beta$ generally increases when the difference in wettability of the two fluids is larger, and when this difference is present for a larger fraction of the porous network.
This is because a larger difference in wettability, meaning that the wetting angle is further away from $90^\circ$, gives rise to a larger interfacial capillary pressure and the overall threshold pressure.
This in turn makes the effect of opening new pathways more prominent.

\section{Declaration}%
\label{sec:declaration}

This version of the article has been accepted for publication in Transport in Porous Media, after peer review but is not the Version of Record and does not reflect post-acceptance improvements, or any corrections. The Version of Record is available online at: http://dx.doi.org/10.1007/s11242-021-01674-3.

\noindent \textbf{Author Contributions}:
HF developed the theory, performed the analytical and numerical calculations of the CFB model, contributed to editing of the code of the DPN model, ran the simulations and analyzed the data of the DPN model and wrote the first draft. 
SS suggested the idea of the problem and developed the code for the DPN model. 
SR sketched the initial calculation related to the CFB model and helped in data analysis.
All the authors contributed in developing the theory and writing the manuscript to its final form.

\noindent \textbf{Funding}: 
This work was partly supported by the Research Council of Norway through its Center of Excellence funding scheme, project number 262644.
SS was partially supported by the National Natural Science Foundation of China under grant number 11750110430. 

\noindent \textbf{Conflicts of interest}:
The authors declare no conflict of interest.



\bibliographystyle{abbrvnat}

\bibliography{bibliography}   

\begin{thebibliography}{36}
\providecommand{\natexlab}[1]{#1}
\providecommand{\url}[1]{\texttt{#1}}
\expandafter\ifx\csname urlstyle\endcsname\relax
  \providecommand{\doi}[1]{doi: #1}\else
  \providecommand{\doi}{doi: \begingroup \urlstyle{rm}\Url}\fi

\bibitem[Aker et~al.(1998)Aker, M{\aa}l{\o}y, Hansen, and Batrouni]{amhb98}
E.~Aker, K.~J. M{\aa}l{\o}y, A.~Hansen, and G.~G. Batrouni.
\newblock A two-dimensional network simulator for two-phase flow in porous
  media.
\newblock \emph{Transport in porous media}, 32\penalty0 (2):\penalty0 163--186,
  1998.
\newblock \doi{https://doi.org/10.1023/A:1006510106194}.

\bibitem[Alhammadi et~al.(2017)Alhammadi, AlRatrout, Singh, Bijeljic, and
  Blunt]{alhammadi2017situ}
A.~M. Alhammadi, A.~AlRatrout, K.~Singh, B.~Bijeljic, and M.~J. Blunt.
\newblock In situ characterization of mixed-wettability in a reservoir rock at
  subsurface conditions.
\newblock \emph{Scientific Reports}, 7\penalty0 (1):\penalty0 1--9, 2017.
\newblock \doi{https://doi.org/10.1038/s41598-017-10992-w}.

\bibitem[AlRatrout et~al.(2018)AlRatrout, Blunt, and
  Bijeljic]{alratrout2018wettability}
A.~AlRatrout, M.~J. Blunt, and B.~Bijeljic.
\newblock Wettability in complex porous materials, the mixed-wet state, and its
  relationship to surface roughness.
\newblock \emph{Proceedings of the National Academy of Sciences}, 115\penalty0
  (36):\penalty0 8901--8906, 2018.
\newblock \doi{https://doi.org/10.1073/pnas.1803734115}.

\bibitem[Anderson(1986)]{anderson1986wettability}
W.~G. Anderson.
\newblock Wettability literature survey-part 1: rock/oil/brine interactions and
  the effects of core handling on wettability.
\newblock \emph{Journal of petroleum technology}, 38\penalty0 (10):\penalty0
  1125--1144, 1986.
\newblock \doi{https://doi.org/10.2118/13932-PA}.

\bibitem[Aursj{\o} et~al.(2014)Aursj{\o}, Erpelding, Tallakstad, Flekk{\o}y,
  Hansen, and M{\aa}l{\o}y]{aursjo2014film}
O.~Aursj{\o}, M.~Erpelding, K.~T. Tallakstad, E.~G. Flekk{\o}y, A.~Hansen, and
  K.~J. M{\aa}l{\o}y.
\newblock Film flow dominated simultaneous flow of two viscous incompressible
  fluids through a porous medium.
\newblock \emph{Frontiers in physics}, 2:\penalty0 63, 2014.
\newblock \doi{https://doi.org/10.3389/fphy.2014.00063}.

\bibitem[Blunt(2017)]{blunt2017multiphase}
M.~J. Blunt.
\newblock \emph{Multiphase flow in permeable media: A pore-scale perspective}.
\newblock Cambridge University Press, 2017.
\newblock \doi{https://doi.org/10.1017/9781316145098}.

\bibitem[Elkhyat et~al.(2001)Elkhyat, Agache, Zahouani, and
  Humbert]{elkhyat2001new}
A.~Elkhyat, P.~Agache, H.~Zahouani, and P.~Humbert.
\newblock A new method to measure in vivo human skin hydrophobia.
\newblock \emph{International journal of cosmetic science}, 23\penalty0
  (6):\penalty0 347--352, 2001.
\newblock \doi{https://doi.org/10.1046/j.0412-5463.2001.00108.x}.

\bibitem[Flovik et~al.(2015)Flovik, Sinha, and Hansen]{flovikSinha2015dynamic}
V.~Flovik, S.~Sinha, and A.~Hansen.
\newblock Dynamic wettability alteration in immiscible two-phase flow in porous
  media: Effect on transport properties and critical slowing down.
\newblock \emph{Frontiers in Physics}, 3:\penalty0 86, 2015.
\newblock \doi{https://doi.org/10.3389/fphy.2015.00086}.

\bibitem[Frette et~al.(1997)Frette, Måløy, Schmittbuhl, and
  Hansen]{frette1997immiscible}
O.~I. Frette, K.~J. Måløy, J.~Schmittbuhl, and A.~Hansen.
\newblock Immiscible displacement of viscosity-matched fluids in
  two-dimensional porous media.
\newblock \emph{Physical Review E}, 55\penalty0 (3):\penalty0 2969, 1997.
\newblock \doi{https://doi.org/10.1103/PhysRevE.55.2969}.

\bibitem[Gao et~al.(2020{\natexlab{a}})Gao, Lin, Bijeljic, and
  Blunt]{gaolin2020pore}
Y.~Gao, Q.~Lin, B.~Bijeljic, and M.~J. Blunt.
\newblock Pore-scale dynamics and the multiphase darcy law.
\newblock \emph{Physical Review Fluids}, 5\penalty0 (1):\penalty0 013801,
  2020{\natexlab{a}}.
\newblock \doi{https://doi.org/10.1103/PhysRevFluids.5.013801}.

\bibitem[Gao et~al.(2020{\natexlab{b}})Gao, Raeini, Selem, Bondino, Blunt, and
  Bijeljic]{gao2020pore}
Y.~Gao, A.~Q. Raeini, A.~M. Selem, I.~Bondino, M.~J. Blunt, and B.~Bijeljic.
\newblock Pore-scale imaging with measurement of relative permeability and
  capillary pressure on the same reservoir sandstone sample under water-wet and
  mixed-wet conditions.
\newblock \emph{Advances in Water Resources}, 146:\penalty0 103786,
  2020{\natexlab{b}}.
\newblock \doi{https://doi.org/10.1016/j.advwatres.2020.103786}.

\bibitem[Gjennestad et~al.(2018)Gjennestad, Vassvik, Kjelstrup, and
  Hansen]{gvkh18}
M.~A. Gjennestad, M.~Vassvik, S.~Kjelstrup, and A.~Hansen.
\newblock Stable and efficient time integration of a dynamic pore network model
  for two-phase flow in porous media.
\newblock \emph{Frontiers in Physics}, 6:\penalty0 56, 2018.
\newblock \doi{https://doi.org/10.3389/fphy.2018.0005}.

\bibitem[Knudsen et~al.(2002)Knudsen, Aker, and Hansen]{kah02}
H.~A. Knudsen, E.~Aker, and A.~Hansen.
\newblock Bulk flow regimes and fractional flow in 2d porous media by numerical
  simulations.
\newblock \emph{Transport in Porous Media}, 47:\penalty0 99--121, 2002.
\newblock \doi{https://doi.org/10.1023/A:1015039503551}.

\bibitem[Kovscek et~al.(1993)Kovscek, Wong, and Radke]{kovscek1993pore}
A.~Kovscek, H.~Wong, and C.~Radke.
\newblock A pore-level scenario for the development of mixed wettability in oil
  reservoirs.
\newblock \emph{AIChE Journal}, 39\penalty0 (6):\penalty0 1072--1085, 1993.
\newblock \doi{https://doi.org/10.1002/aic.690390616}.

\bibitem[Krevor et~al.(2015)Krevor, Blunt, Benson, Pentland, Reynolds,
  Al-Menhali, and Niu]{KREVOR2015221}
S.~Krevor, M.~J. Blunt, S.~M. Benson, C.~H. Pentland, C.~Reynolds,
  A.~Al-Menhali, and B.~Niu.
\newblock Capillary trapping for geologic carbon dioxide storage – from pore
  scale physics to field scale implications.
\newblock \emph{International Journal of Greenhouse Gas Control}, 40:\penalty0
  221--237, 2015.
\newblock \doi{https://doi.org/10.1016/j.ijggc.2015.04.006}.

\bibitem[Lenormand and Zarcone(1989)]{lenormand1989capillary}
R.~Lenormand and C.~Zarcone.
\newblock Capillary fingering: percolation and fractal dimension.
\newblock \emph{Transport in porous media}, 4\penalty0 (6):\penalty0 599--612,
  1989.
\newblock \doi{https://doi.org/10.1007/BF00223630}.

\bibitem[Li et~al.(2017)Li, Huang, Chen, Chen, and Lai]{li2017review}
S.~Li, J.~Huang, Z.~Chen, G.~Chen, and Y.~Lai.
\newblock A review on special wettability textiles: theoretical models,
  fabrication technologies and multifunctional applications.
\newblock \emph{Journal of Materials Chemistry A}, 5\penalty0 (1):\penalty0
  31--55, 2017.
\newblock \doi{https://doi.org/10.1039/c6ta07984a}.

\bibitem[L{\o}voll et~al.(2004)L{\o}voll, M{\'e}heust, Toussaint, Schmittbuhl,
  and M{\aa}l{\o}y]{lovoll2004growth}
G.~L{\o}voll, Y.~M{\'e}heust, R.~Toussaint, J.~Schmittbuhl, and K.~J.
  M{\aa}l{\o}y.
\newblock Growth activity during fingering in a porous hele-shaw cell.
\newblock \emph{Physical Review E}, 70\penalty0 (2):\penalty0 026301, 2004.
\newblock \doi{https://doi.org/10.1103/PhysRevE.70.026301}.

\bibitem[M{\aa}l{\o}y et~al.(1985)M{\aa}l{\o}y, Feder, and
  J{\o}ssang]{maaloy1985viscous}
K.~J. M{\aa}l{\o}y, J.~Feder, and T.~J{\o}ssang.
\newblock Viscous fingering fractals in porous media.
\newblock \emph{Physical review letters}, 55\penalty0 (24):\penalty0 2688,
  1985.
\newblock \doi{https://doi.org/10.1103/PhysRevLett.55.2688}.

\bibitem[Marle(1981)]{marle1981multiphase}
C.~Marle.
\newblock \emph{Multiphase flow in porous media}.
\newblock {\'E}ditions technip, 1981.

\bibitem[M{\'e}heust et~al.(2002)M{\'e}heust, L{\o}voll, M{\aa}l{\o}y, and
  Schmittbuhl]{meheust2002interface}
Y.~M{\'e}heust, G.~L{\o}voll, K.~J. M{\aa}l{\o}y, and J.~Schmittbuhl.
\newblock Interface scaling in a two-dimensional porous medium under combined
  viscous, gravity, and capillary effects.
\newblock \emph{Physical Review E}, 66\penalty0 (5):\penalty0 051603, 2002.
\newblock \doi{https://doi.org/10.1103/PhysRevE.66.051603}.

\bibitem[Rassi et~al.(2011)Rassi, Codd, and Seymour]{rassi2011nuclear}
E.~M. Rassi, S.~L. Codd, and J.~D. Seymour.
\newblock Nuclear magnetic resonance characterization of the stationary
  dynamics of partially saturated media during steady-state infiltration flow.
\newblock \emph{New Journal of Physics}, 13\penalty0 (1):\penalty0 015007,
  2011.
\newblock \doi{https://doi.org/10.1088/1367-2630/13/1/015007}.

\bibitem[Roy et~al.(2019)Roy, Hansen, and Sinha]{roy2019effective}
S.~Roy, A.~Hansen, and S.~Sinha.
\newblock Effective rheology of two-phase flow in a capillary fiber bundle
  model.
\newblock \emph{Frontiers in Physics}, 2019.
\newblock \doi{https://doi.org/10.3389/fphy.2019.00092}.

\bibitem[Salathiel(1973)]{salathiel1972oil}
R.~Salathiel.
\newblock Oil recovery by surface film drainage in mixed-wettability rocks.
\newblock \emph{Journal of petroleum technology}, 25\penalty0 (10):\penalty0
  1216--1224, 10 1973.
\newblock \doi{https://doi.org/10.2118/4104-PA}.

\bibitem[Scanziani et~al.(2020)Scanziani, Lin, Alhosani, Blunt, and
  Bijeljic]{scanziani2020dynamics}
A.~Scanziani, Q.~Lin, A.~Alhosani, M.~J. Blunt, and B.~Bijeljic.
\newblock Dynamics of fluid displacement in mixed-wet porous media.
\newblock \emph{Proceedings of the Royal Society A}, 476\penalty0
  (2240):\penalty0 20200040, 2020.
\newblock \doi{https://doi.org/10.1098/rspa.2020.0040}.

\bibitem[Sinha and Hansen(2012)]{sinha2012effective}
S.~Sinha and A.~Hansen.
\newblock Effective rheology of immiscible two-phase flow in porous media.
\newblock \emph{EPL (Europhysics Letters)}, 99\penalty0 (4):\penalty0 44004,
  2012.
\newblock \doi{https://doi.org/10.1209/0295-5075/99/44004}.

\bibitem[Sinha et~al.(2011)Sinha, Gr{\o}va, {\O}deg{\aa}rden, Skjetne, and
  Hansen]{sinhaGrova2011local}
S.~Sinha, M.~Gr{\o}va, T.~B. {\O}deg{\aa}rden, E.~Skjetne, and A.~Hansen.
\newblock Local wettability reversal during steady-state two-phase flow in
  porous media.
\newblock \emph{Physical Review E}, 84\penalty0 (3):\penalty0 037303, 2011.
\newblock \doi{https://doi.org/10.1103/PhysRevE.84.037303}.

\bibitem[Sinha et~al.(2013)Sinha, Hansen, Bedeaux, and
  Kjelstrup]{sinha2013effective}
S.~Sinha, A.~Hansen, D.~Bedeaux, and S.~Kjelstrup.
\newblock Effective rheology of bubbles moving in a capillary tube.
\newblock \emph{Physical Review E}, 87\penalty0 (2):\penalty0 025001, 2013.
\newblock \doi{https://doi.org/10.1103/PhysRevE.87.025001}.

\bibitem[Sinha et~al.(2021)Sinha, Gjennestad, Vassvik, and
  Hansen]{sinha2019dynamic}
S.~Sinha, M.~A. Gjennestad, M.~Vassvik, and A.~Hansen.
\newblock Fluid meniscus algorithms for dynamic pore-network modeling of
  immiscible two-phase flow in porous media.
\newblock \emph{Frontiers in Physics}, 8:\penalty0 567, 2021.
\newblock ISSN 2296-424X.
\newblock \doi{https://doi.org/10.3389/fphy.2020.548497}.

\bibitem[Tallakstad et~al.(2009{\natexlab{a}})Tallakstad, Knudsen, Ramstad,
  L{\o}voll, M{\aa}l{\o}y, Toussaint, and Flekk{\o}y]{tallakstad2009steady}
K.~T. Tallakstad, H.~A. Knudsen, T.~Ramstad, G.~L{\o}voll, K.~J. M{\aa}l{\o}y,
  R.~Toussaint, and E.~G. Flekk{\o}y.
\newblock Steady-state two-phase flow in porous media: statistics and transport
  properties.
\newblock \emph{Physical review letters}, 102\penalty0 (7):\penalty0 074502,
  2009{\natexlab{a}}.
\newblock \doi{https://doi.org/10.1103/PhysRevLett.102.074502}.

\bibitem[Tallakstad et~al.(2009{\natexlab{b}})Tallakstad, L{\o}voll, Knudsen,
  Ramstad, Flekk{\o}y, and M{\aa}l{\o}y]{tallakstad2009steadyE}
K.~T. Tallakstad, G.~L{\o}voll, H.~A. Knudsen, T.~Ramstad, E.~G. Flekk{\o}y,
  and K.~J. M{\aa}l{\o}y.
\newblock Steady-state, simultaneous two-phase flow in porous media: An
  experimental study.
\newblock \emph{Physical Review E}, 80\penalty0 (3):\penalty0 036308,
  2009{\natexlab{b}}.
\newblock \doi{https://doi.org/10.1103/PhysRevE.80.036308}.

\bibitem[T{\o}r{\aa} et~al.(2012)T{\o}r{\aa}, {\O}ren, and Hansen]{toh12}
G.~T{\o}r{\aa}, P.-E. {\O}ren, and A.~Hansen.
\newblock A dynamic network model for two-phase flow in porous media.
\newblock \emph{Transport in Porous Media}, 92\penalty0 (1):\penalty0 145--164,
  2012.
\newblock \doi{https://doi.org/10.1007/s11242-011-9895-6}.

\bibitem[Toussaint et~al.(2005)Toussaint, L{\o}voll, M{\'e}heust, M{\aa}l{\o}y,
  and Schmittbuhl]{toussaint2005influence}
R.~Toussaint, G.~L{\o}voll, Y.~M{\'e}heust, K.~J. M{\aa}l{\o}y, and
  J.~Schmittbuhl.
\newblock Influence of pore-scale disorder on viscous fingering during
  drainage.
\newblock \emph{EPL (Europhysics Letters)}, 71\penalty0 (4):\penalty0 583,
  2005.
\newblock \doi{https://doi.org/10.1209/epl/i2005-10136-9}.

\bibitem[Vafai(2010)]{vafai2010porous}
K.~Vafai.
\newblock \emph{Porous media: applications in biological systems and
  biotechnology}.
\newblock CRC Press, 2010.
\newblock \doi{https://doi.org/10.1201/9781420065428}.

\bibitem[Washburn(1921)]{washburn1921dynamics}
E.~W. Washburn.
\newblock The dynamics of capillary flow.
\newblock \emph{Physical review}, 17\penalty0 (3):\penalty0 273, 1921.
\newblock \doi{https://doi.org/10.1103/PhysRev.17.273}.

\bibitem[Zhang et~al.(2021)Zhang, Bijeljic, Gao, Lin, and
  Blunt]{zhang2021quantification}
Y.~Zhang, B.~Bijeljic, Y.~Gao, Q.~Lin, and M.~J. Blunt.
\newblock Quantification of nonlinear multiphase flow in porous media.
\newblock \emph{Geophysical Research Letters}, 48\penalty0 (5):\penalty0
  e2020GL090477, 2021.
\newblock \doi{https://doi.org/10.1029/2020GL090477}.

\end{thebibliography}


\end{document}